\documentclass[11pt]{article}
\def\mytitle#1{\setcounter{equation}{0}
\setcounter{footnote}{0}
\begin{center}\Large\textbf{#1}\end{center}
\vspace{0.25cm}}
\def\myname#1{\begin{center}{{\large #1}}\end{center}\vspace{-0.13cm}}
\def\myplace#1#2{\small\begin{center}\textit{#1}\\
\texttt{#2}\end{center}}

\def\myclassification#1{\small\indent
keywords:
#1\vspace{0.5cm}}
\usepackage{amsmath}
\usepackage{ragged2e}
\usepackage{amssymb}
\usepackage[section]{placeins}
\usepackage{graphicx}
\usepackage[utf8]{inputenc}
\usepackage{mathrsfs}
\usepackage{float}
\usepackage{hyperref}
\usepackage{xcolor}
\begin{document}
\mytitle{Stability, Tidal Deformability, and Statistical Constraints of Anisotropic Compact Stars with a Generalized Polytropic Equation of State : Modelling Observation with Buchdahl Space-Time}

\myname{$Ritabrata~Biswas^{\dag}$\footnote{biswas.ritabrata@gmail.com \bf{(ORCID: 0000-0003-3086-892X)} [corresponding author]
}, $Arpan~Bhattacharya^{\dag}$\footnote{arpanbhattacharya616@gmail.com \bf{(ORCID: 0009-0008-1820-1464)}} and  Subhajit Pal $^*$\footnote{subhajitpal968@gmail.com \bf{(ORCID: 0009-0002-2431-1375)}}}
\myplace{${}^{\dag}$Department of Mathematics, The University of Burdwan, Purba Bardhaman-$713104$, West Bengal, India.\\$^*$Department of Mathematics, Jadavpur University, Kolkata-32, India.} {}

\begin{abstract}

An anisotropic compact star model is constructed within the framework of general relativity by combining the Buchdahl space-time with a generalized polytropic equation of state. The model parameters are constrained through Darmois-Israel junction conditions by matching interior geometry smoothly with exterior Schwarzschild space-time. The proposed framework is applied to thirty-eight observed compact star candidates spanning a broad range of masses and radii. Regular and physically acceptable distributions of the energy density, radial pressure, tangential pressure, enclosed mass, anisotropy and gravitational redshift are obtained throughout the stellar interior. Energy density and pressure components are found to decrease outward, while the enclosed mass and gravitational redshift are found to increase monotonically towards the stellar surface. The radial pressure is required to vanish at the boundary, thereby determining the estimated stellar radius. The anisotropy is found to provide an additional outward contribution that supports the stellar configurations against gravitational collapse. The physical acceptability of the model is examined through the Zeldovich condition, radial adiabatic index, causality conditions and Herrera cracking criterion. Hydrostatic equilibrium is achieved through the balance among gravitational, hydrostatic and anisotropic forces. The tidal response is also investigated, and an inverse relation between the estimated mass and dimensionless tidal deformability is obtained. The calculated configurations span \(2.88\lesssim M_{\rm Est}/M_\odot\lesssim3.23\) and yield \(22.55\lesssim\Lambda\lesssim68.38\). MCMC analysis provides statistically constrained compactness and tidal parameters. Overall, a physically viable description of anisotropic compact stars is obtained.

\end{abstract}  

\myclassification{Neutron stars, Stellar structure, General relativity.}
\section{Introduction}
Compact stars, particularly neutron stars, are among the densest astrophysical objects in the Universe and provide a unique laboratory for investigating the properties of matter at supranuclear densities within the framework of general relativity (GR) \cite{Tolman1939,Oppenheimer1939,Glendenning2000,Haensel2007,Lattimer2016}. Recent observations from Neutron star Interior Composition ExploreR (NICER) and gravitational wave (GW) events have imposed stringent constraints on the mass--radius relation and the equation of state (EoS) of neutron stars \cite{Abbott2017,Riley2019,Miller2019,Riley2021,Miller2021}. Although the earliest stellar models assumed isotropic matter distributions, Bowers and Liang demonstrated that pressure anisotropy can significantly modify the equilibrium configuration, maximum mass, and surface redshift of relativistic stars \cite{Bowers1974}. In anisotropic compact stars, the radial thermal pressure balances the inward gravitational pull, whereas the tangential pressure produces an additional anisotropic force that can support more massive and compact stellar configurations \cite{Herrera1997,Dev2002,Mak2002}. Consequently, numerous anisotropic solutions based on different metric potentials and EoSs have been proposed to describe realistic neutron stars while satisfying all physical acceptability conditions \cite{Maurya2017,Sharma2020,Deb2021}. Motivated by these developments, the present work investigates anisotropic neutron stars in Buchdahl space-time using a generalized polytropic EoS and examines their physical properties in the light of current observational constraints.

The uncertain nature of the EoS of matter at supranuclear densities continues to present a major challenge in the theoretical modelling of compact stars, thereby motivating the development of phenomenological models capable of describing the complex behaviour of ultra-dense-matter within the framework of GR \cite{Glendenning2000,Haensel2007,Lattimer2016}. Recent high-precision measurements of neutron star masses, radii, and tidal deformabilities from radio timing, NICER observations, and GW detections have imposed stringent observational constraints on viable relativistic stellar models, requiring them to satisfy both astrophysical measurements and fundamental physical acceptability conditions \cite{Abbott2017,Riley2019,Miller2019,Riley2021,Miller2021,Cromartie2020}. Motivated by these developments, the present work constructs an anisotropic compact star model by combining the Buchdahl interior space-time with a generalized polytropic EoS, whose nonlinear terms effectively describe the complex interactions expected in dense nuclear matter \cite{Lattimer2020,Chavanis2015}. The unknown model parameters are determined through the junction conditions together with observational data of realistic compact stars, thereby establishing an observation-constrained relativistic stellar model. The proposed framework is subsequently employed to investigate the radial behaviour of the energy density, pressures, anisotropy, enclosed mass, compactness, surface redshift, causality, equilibrium, and stability of the resulting stellar configurations. Finally, the theoretical predictions are systematically compared with the observed properties of several well-studied compact stars to assess the physical viability and astrophysical applicability of the model.

Although numerous anisotropic models of compact stars have been proposed, a physically acceptable relativistic solution capable of describing a large population of observed neutron stars with a realistic EoS remains desirable. The uncertainty in the behaviour of matter at supranuclear densities further motivates the exploration of alternative interior space-time geometries and EoSs that satisfy all regularity, causality, stability, and matching conditions. In particular, the Buchdahl space-time provides a singularity-free interior geometry, while the generalized polytropic EoS offers a more flexible description of dense-matter than conventional linear models. Motivated by recent observational constraints from massive pulsars and GW sources, the present work investigates whether this combined framework can consistently reproduce the structural and physical characteristics of a wide range of observed compact stars.

The remainder of this article is organized as follows. Section 2 presents the theoretical framework for modelling anisotropic compact stars, where the Einstein field equations are formulated using the Buchdahl space-time and the generalized polytropic EoS, followed by the derivation of the corresponding stellar structure equations and the matching conditions. Section 3 introduces the observational compact star dataset employed in the present analysis and describes the determination of the model constants from the observed mass--radius measurements. In Section 4, the physical behaviour of the resulting stellar configurations is investigated through a comprehensive analysis of the energy density, radial and tangential pressures, anisotropy, mass distribution, compactness, surface redshift, causality, stability, and equilibrium, and the theoretical predictions are compared with the available observational constraints. Finally, the principal findings of the present investigation, together with their astrophysical implications and concluding remarks, are summarized in Section 5.
\section{Framework for Anisotropic Compact Star Modelling : Einstein Field Equations and Stellar Structure}
The energy-momentum tensor for a locally anisotropic fluid, in which the radial and
tangential pressures are treated as independent quantities, was first introduced in
the seminal work of Bowers and Liang \cite{BowersLiang1974}, who demonstrated that
locally anisotropic EoSs can significantly modify the equilibrium
structure, maximum mass, and surface redshift of relativistic spheres relative to the
isotropic case. Following this formalism, the stress-energy tensor for the present
model is written as \cite{Dev2002,Mak2002}
\begin{equation} \label{eq1}
T_{\mu\nu} = (\rho+ p_t)u_{\mu}u_{\nu} + p_t g_{\mu\nu} + (p_{r}-p_t)v_{\mu}v_{\nu}~~~,~~~u^{\mu} = e^{-\frac{\alpha(r)}{2}}\delta_{0}^{\mu}~,~v^{\mu} = e^{-\frac{\beta(r)}{2}}\delta_{1}^{\mu}~~~,
\end{equation}
where $u^\mu$ and $v^\mu$ denote the fluid four-velocity and the unit space-like vector
along the radial direction, respectively. Throughout this work, we adopt geometrized units, in which the Newtonian gravitational constant and the speed of light are set equal to unity, i.e., $G=c=1$. This stress-energy tensor is coupled to the space-time geometry through the standard Einstein field equations of GR \cite{Tolman1939},
\begin{equation}\label{eq2}
G_{\mu\nu}=R_{\mu\nu}-\frac{1}{2}Rg_{\mu\nu}=8\pi T_{\mu\nu}~~~~.
\end{equation}

To close this system, a relation between the radial pressure and the energy density is
required. Polytropic EoSs have long been used to model the interior
structure of both isotropic and anisotropic relativistic stars, since they provide
analytically tractable solutions while capturing the essential density dependence of
the pressure expected in dense stellar matter. In particular, Thirukkanesh and Ragel
\cite{ThirukkaneshRagel2012} obtained exact anisotropic sphere solutions using a
polytropic EoS for several choices of the polytropic index, showing that
the choice $n=2$ yields mass and density profiles consistent with observed compact
objects, while Mafa Takisa and Maharaj \cite{TakisaMaharaj2013} extended this approach
to charged anisotropic configurations with a general polytropic EoS.

Motivated by these results, the present model adopts a generalized (modified) polytropic EoS that relates the radial pressure to the energy density in the form \cite{ThirukkaneshRagel2012,TakisaMaharaj2013},
\begin{equation}\label{eq:eos}
p_r = X\rho^{1+\frac{1}{n}} - Y ~~, \qquad 0.5 \leq n \leq 3 ~,
\end{equation}
where $n$ is the polytropic index and $X$, $Y$ are constants fixed, together with the
other model parameters, by the physical requirements $p_r \geq 0$, $\rho > 0$, and
$0 \leq v_r^2 \leq 1$ throughout the stellar interior. Here, $v_r^{2}=dp_r/d\rho$ denotes the squared radial sound speed within the stellar interior, which characterizes the propagation of infinitesimal radial pressure perturbations through the stellar matter. In this regard we take $n=2$, so that
Eq.~(\ref{eq:eos}) reduces to \cite{Kumara2025}
\begin{equation}\label{eq:eos_n2}
p_r = X\rho^{\frac{3}{2}} - Y ~~.
\end{equation}
Here, $X$ and $Y$ denote the EoS parameters and are chosen such that the fundamental physical acceptability conditions, namely $p_r\geq0$, $\rho>0$, and $0\leq v_r^{2}\leq1$, are satisfied throughout the stellar interior. The first two conditions ensure the positivity of both the pressure and the energy density, whereas the last condition guarantees that the propagation of pressure perturbations remains causal. Consequently, the adopted generalized polytropic EoS remains thermodynamically consistent and physically admissible over the entire stellar configuration \cite{Herrera1997,Canuto1975,Tooper1965,Delgaty1998}. Throughout the analysis, we adopt geometrized units for the EoS parameters and fix them at $X=5\,\mathrm{km}$ and $Y=1.3\times10^{-7}\,\mathrm{km}^{-2}$.

\vspace{0.2cm}

To model the interior gravitational field of a static compact object, we consider the most general spherically symmetric line element expressed in Schwarzschild-like coordinates as
\begin{equation}\label{eq3}
ds_{-}^{2}
=
-e^{\alpha(r)}dt^{2}
+
e^{\beta(r)}dr^{2}
+
r^{2}
\left(
d\theta^{2}
+\sin^{2}\theta\,d\phi^{2}
\right),
\end{equation}
where $\alpha(r)$ and $\beta(r)$ are the metric potentials depending only on the radial coordinate $r$. In this regard $0\leq\theta\leq\pi ~,~ 0\leq\phi\leq2\pi$. This metric provides an appropriate description of static, self-gravitating relativistic fluid spheres and has been extensively employed in the modelling of neutron stars and other compact astrophysical objects \cite{Tolman1939,Oppenheimer1939,Delgaty1998,Schwarzschild1916}.

Following the Buchdahl prescription for constructing regular interior solutions of Einstein's field equations, we adopt the metric potential \cite{Tolman1939,Buchdahl1959,Amin2026}
\[
e^{-\beta(r)}
=
\frac{(1+ar^{2})(1-cr^{2})}{1+2ar^{2}},
\]
where $a$ and $c$ are positive geometric parameters characterizing the interior space-time. This choice is free from central singularities, remains finite throughout the stellar interior, and has been shown to generate physically acceptable compact star models satisfying the standard regularity conditions \cite{Maurya2017,Buchdahl1959,Sharma2007}.

Substituting the above ansatz together with the generalized polytropic EoS into the Einstein field equations yields the remaining metric potential in the following closed analytical form
\[
\alpha(r)=
\frac{1}{32\pi}
\Bigg\{
-\frac{
2 \log \left(a r^2+1\right)
\left(
a^2 X+4acX-8\pi a+4c^2X-64\pi^2Y
\right)
}{a+c}
\]

\[
-\frac{
2 \log \left(1-cr^2\right)
\left[
16a^2\left(c^2X+2\pi c-16\pi^2Y\right)
+8ac\left(3c^2X+4\pi c-32\pi^2Y\right)
+c^2\left(9c^2X+8\pi c-64\pi^2Y\right)
\right]
}{
c\left(2a^2+3ac+c^2\right)
}
\]

\[
+\frac{
X\left(4a^2+28ac+17c^2\right)
\log\left(2ar^2+1\right)
}{2a+c}
-\frac{4cX}{2ar^2+1}
-\frac{2X(2a+c)}{(2ar^2+1)^2}
\Bigg\}
+A.
\]

The integration constant $A$, together with the geometric parameters $a$ and $c$, completely specifies the interior space-time. Their numerical values are determined by imposing the Darmois--Israel junction conditions at the stellar boundary, which require the continuity of the first and second fundamental forms across the matching hypersurface. Specifically, the interior metric is smoothly joined to the exterior Schwarzschild space-time through the continuity of the metric coefficients, while the radial pressure is required to vanish at the stellar surface. These conditions uniquely determine the unknown constants for each observed compact star candidate and ensure a physically consistent global solution \cite{Schwarzschild1916,Darmois1927,Israel1966}.

\vspace{0.2cm}

Substituting the interior metric (\ref{eq3}) together with the anisotropic energy--momentum tensor (\ref{eq1}) into Einstein's field Eqs. \ref{eq2}, yields the following independent differential equations governing the stellar structure \cite{Tolman1939,Oppenheimer1939,Einstein1916}

\begin{equation}\label{eq4}
1-e^{-\beta}+r\beta'e^{-\beta}
=
8\pi r^2\rho(r),
\end{equation}

\begin{equation}\label{eq5}
e^{-\beta}-1+r\alpha'e^{-\beta}
=
8\pi r^2p_r(r)
\qquad\text{and}
\end{equation}

\begin{equation}\label{eq6}
\frac14re^{-\beta}
\left\{
r(\alpha')^2
-2\beta'
+\alpha'(2-r\beta')
+2r\alpha''
\right\}
=
8\pi r^2p_t(r).
\end{equation}

Here, $\rho$, $p_r$, and $p_t$ denote the energy density, radial pressure, and tangential pressure, respectively, while the prime ($'$) represents differentiation with respect to the radial coordinate $r$. Consequently, all physical quantities are expressed in geometric units unless stated otherwise. Solving Eq.~(\ref{eq5}) for the derivative of the temporal metric potential gives

\[
\frac{d\alpha(r)}{dr}
=
e^{\beta(r)}
\bigg\{
8\pi rp_r
+\frac{1-e^{-\beta(r)}}{r}
\bigg\},
\]
which plays an important role in determining the hydrostatic equilibrium of the stellar configuration.

\vspace{0.2cm}

The equilibrium of a static anisotropic compact star is governed by the conservation equation
$\nabla_{\mu}T^{\mu\nu}=0$, which leads to the generalized Tolman--Oppenheimer--Volkoff (TOV) equation. Together with the enclosed gravitational mass function, the stellar structure equations take the form \cite{Tolman1939,Oppenheimer1939,Bowers1974,Herrera1997}

\begin{equation}\label{eq7}
\frac{dM(r)}{dr}
=
4\pi r^2\rho(r),
\qquad\text{and}
\end{equation}

\begin{equation}\label{eq8}
\frac{dp_r}{dr}
=
-\frac12
\bigg\{
\rho(r)+p_r(r)
\bigg\}
\frac{d\alpha(r)}{dr}
+
\frac{2}{r}
\bigg\{
p_t(r)-p_r(r)
\bigg\}.
\end{equation}

The first equation describes the accumulation of gravitational mass from the stellar centre to the boundary, whereas the second expresses the balance among the inward gravitational force, the outward hydrostatic pressure gradient, and the force arising from local pressure anisotropy. The stellar surface is defined by the vanishing of the radial pressure,

\[
p_r(R)=0,
\]
which determines the radius of the compact object \cite{Bowers1974,Herrera1997}. Here, $M(r)$ denotes the total mass enclosed within the stellar configuration up to the radial coordinate $r$, i.e., the cumulative mass contained inside a sphere of radius $r$.

\vspace{0.2cm}

Outside the stellar boundary $(r\geq R)$, both the energy density and pressure vanish identically, and the interior solution must be matched smoothly to the Schwarzschild vacuum space-time. Accordingly, the exterior line element is given by \cite{Schwarzschild1916,Darmois1927,Israel1966}

\begin{equation}\label{eq9}
ds_+^{2}
=
-\left\{1-\frac{2M(R)}{r}\right\}dt^2
+\left\{1-\frac{2M(R)}{r}\right\}^{-1}dr^2
+r^2\left(d\theta^2
+\sin^2\theta\,d\phi^2\right).
\end{equation}

The unknown constants appearing in the interior solution are determined by imposing the Darmois--Israel junction conditions at the hypersurface $r=R$, which require the continuity of the metric coefficients together with the vanishing of the radial pressure. These matching conditions ensure a smooth transition between the interior matter distribution and the exterior vacuum geometry, thereby yielding a physically complete relativistic stellar model \cite{Darmois1927,Israel1966}.

\vspace{0.2cm}

The interior solution must be matched smoothly to the exterior Schwarzschild vacuum space-time across the stellar hypersurface $r=R$. According to the Darmois--Israel junction formalism, the first fundamental form must be continuous across the boundary, while the radial pressure must vanish at the stellar surface. These conditions ensure the absence of any surface layer or thin shell and guarantee a physically acceptable transition from the interior matter distribution to the exterior vacuum geometry \cite{Darmois1927,Israel1966}.

Accordingly, the matching conditions are given by
\begin{equation}\label{eq10}
g_{tt}^{-}=g_{tt}^{+}, \qquad
g_{rr}^{-}=g_{rr}^{+}, \qquad
p_r(R)=0.
\end{equation}

Using the interior metric (\ref{eq3}) and the exterior Schwarzschild space-time (\ref{eq9}), the above conditions reduce to

\[
e^{\alpha(R)}
=
1-\frac{2M(R)}{R},
\qquad
e^{\beta(R)}
=
\left\{1-\frac{2M(R)}{R}\right\}^{-1},
\qquad
p_r(R)=0.
\]

The first two equations ensure the continuity of the temporal and radial metric potentials across the boundary, whereas the third condition defines the stellar surface as the radius at which the radial pressure vanishes. Together, these boundary conditions uniquely determine the integration constant $A$ and the geometric parameters $a$ and $c$ for each observed compact star candidate. Consequently, the resulting interior solution is completely specified and joins smoothly to the exterior Schwarzschild space-time without introducing any discontinuity in the space-time geometry \cite{Schwarzschild1916,Darmois1927,Israel1966}. The total gravitational mass enclosed within the stellar radius $R$ is obtained by integrating the mass equation, Eq.~\eqref{eq7}, from the stellar centre to the surface. Consequently, the total gravitational mass of the compact star is given by
\begin{equation}
M \equiv M(R)=4\pi\int_{r=0}^{r=R}\rho(r)\,r^{2}\,dr.
\end{equation}
\section{Observed Compact Star Mass–Radius Data and Derived Model Constants}
\begin{table}
\centering
\small
\caption{\textbf{Model constants of various neutron stars for the constructed stellar model.}}
\label{tab:model_constants}
\vspace{0.1in}
\begin{tabular}{|c|l|c|c|c|}
\hline
\textbf{Sl. No.} & \textbf{Neutron Star} & \textbf{$a$} [$km^{-2}$] & \textbf{$c$} [$km^{-2}$] & \textbf{$A$} [$km^{0}$] \\
\hline
$L_{01}$ & $Her~X\!-\!1$ \cite{Rawls2011}
                 & 0.001834288 & 0.000541159 & 0.980843 \\ \hline
$L_{02}$ & $4U~1538\!-\!52$ \cite{Ozel2009}
             & 0.001840470 & 0.000532748 & 0.983063 \\ \hline
$L_{03}$ & $SAX~J1808.4\!-\!3658$ \cite{Demorest2010}
       & 0.001842479 & 0.000530016 & 0.983789 \\ \hline
$L_{04}$ & $LMC~X\!-\!4$ \cite{Antoniadis2013}
                & 0.001810060 & 0.000574147 & 0.704635 \\ \hline
$L_{05}$ & $Cen~X\!-\!4$ \cite{Maurya2023}
                & 0.001813129 & 0.000569966 & 0.700402 \\ \hline
$L_{06}$ & $SMC~X\!-\!1$ \cite{Maurya2024}
                & 0.001818184 & 0.000563081 & 0.692261 \\ \hline
$L_{07}$ & $GW170817\!-\!2$ \cite{Maurya2024}
            & 0.001821414 & 0.000558682 & 0.697408 \\ \hline
$L_{08}$ & $SMC~X\!-\!4$ \cite{Cromartie2020}
               & 0.001817110 & 0.000564544 & 0.719335 \\ \hline
$L_{09}$ & $EXO~1785\!-\!248$ \cite{Romani2022}
           & 0.001819272 & 0.000561600 & 0.711168 \\ \hline
$L_{10}$ & $PSR~J0030\!+\!0451$ \cite{Maurya2025MassGap}
         & 0.001824959 & 0.000553856 & 0.697074 \\ \hline
$L_{11}$ & $X7~(47~Tuc)$ \cite{Bogdanov2016}
               & 0.001821418 & 0.000558677 & 0.714233 \\ \hline
$L_{12}$ & $Swift~J1818.0\!-\!1607$ \cite{Singh2026}
     & 0.001823434 & 0.000555932 & 0.703922 \\ \hline
$L_{13}$ & $PSR~B0943\!+\!10$ \cite{Maurya2026JCAP}
           & 0.001823762 & 0.000555486 & 0.707878 \\ \hline
$L_{14}$ & $M13~neutron~star$ \cite{Shaw2018}
           & 0.001824389 & 0.000554632 & 0.703135 \\ \hline
$L_{15}$ & $PSR~J0437\!-\!4715$ \cite{Choudhury2024}
         & 0.001820953 & 0.000559310 & 0.711014 \\ \hline
$L_{16}$ & $PSR~J0614\!-\!3329$ \cite{Mauviard2025}
         & 0.001820323 & 0.000560168 & 0.730260 \\ \hline
$L_{17}$ & $Cen~X\!-\!3$ \cite{delaCruzDombriz2009}
               & 0.001820428 & 0.000560025 & 0.735826 \\ \hline
$L_{18}$ & $PSR~B0656\!+\!14$ \cite{Maurya2022JCAP}
           & 0.001828485 & 0.000549056 & 0.708080 \\ \hline
$L_{19}$ & $XTE~J1739\!-\!285$ \cite{Kaaret2007}
          & 0.001823001 & 0.000556522 & 0.729388 \\ \hline
$L_{20}$ & $4U~1724\!-\!307$ \cite{Hansraj2024}
            & 0.001825666 & 0.000552894 & 0.710591 \\ \hline
$L_{21}$ & $4U~1820\!-\!30$ \cite{Kaaret1999}
             & 0.001824883 & 0.000553959 & 0.738124 \\ \hline
$L_{22}$ & $Aql~X\!-\!1$ \cite{Li2017}
                & 0.001824080 & 0.000555052 & 0.754407 \\ \hline
$L_{23}$ & $KS~1731\!-\!260$ \cite{Maurya2025FQT}
            & 0.001825182 & 0.000553553 & 0.735723 \\ \hline
$L_{24}$ & $PSR~J1903\!+\!327$ \cite{Linares2018}
       & 0.001826970 & 0.000551118 & 0.734109 \\ \hline
$L_{25}$ & $Cyg~X\!-\!2$ \cite{Casares2010}
                & 0.001830680 & 0.000546069 & 0.731939 \\ \hline
$L_{26}$ & $4U~1608\!-\!52$ \cite{Freire2011}
             & 0.001829159 & 0.000548139 & 0.751260 \\ \hline
$L_{27}$ & $Vela~X\!-\!1$ \cite{Abubekerov2008}
               & 0.001832487 & 0.000543610 & 0.733732 \\ \hline
$L_{28}$ & $4U~1724\!-\!207$ \cite{Guver2010a}
            & 0.001832812 & 0.000543168 & 0.734307 \\ \hline
$L_{29}$ & $SAX~J1748.9\!-\!2021$ \cite{Guver2010b}
       & 0.001831188 & 0.000545377 & 0.750546 \\ \hline
$L_{30}$ & $4U~1702\!-\!429$ \cite{Singh2022Lovelock}
            & 0.001836246 & 0.000538494 & 0.746992 \\ \hline
$L_{31}$ & $PSR~J1614\!-\!2230$ \cite{Elebert2009}
         & 0.001839319 & 0.000534313 & 0.742077 \\ \hline
$L_{32}$ & $PSR~J0348\!+\!0432$ \cite{Abubekerov2008}
         & 0.001840750 & 0.000532368 & 0.749186 \\ \hline
$L_{33}$ & $PSR~J0740\!+\!6620$ \cite{Linares2018}
         & 0.001841319 & 0.000531594 & 0.768814 \\ \hline
$L_{34}$ & $PSR~J1748\!-\!2446ad$ \cite{Elebert2009}
       & 0.001841920 & 0.000530776 & 0.767221 \\ \hline
$L_{35}$ & $PSR~J2215\!+\!5135$ \cite{delaCruzDombriz2009}
         & 0.001847335 & 0.000523412 & 0.780678 \\ \hline
$L_{36}$ & $PSR~J0952\!-\!0607$ \cite{Rawls2011}
         & 0.001922057 & 0.000421998 & 0.998412 \\ \hline
$L_{37}$ & $4U~1700\!-\!37$ \cite{Hansraj2025}
             & 0.001850184 & 0.000519537 & 0.870521 \\ \hline
$L_{38}$ & $GW190814\!-\!2$ \cite{Maurya2024ApJ}
             & 0.001928491 & 0.000413283 & 0.996171 \\ \hline
\end{tabular}
\vspace{2mm}
\footnotesize
\begin{justify}
\textit{\bf{Table~1:}} lists the model constants, $a$, $c$, and $A$ obtained for each of the thirty-eight compact objects considered in this work. These constants appear in the interior metric functions $\alpha(r)$ and $\beta(r)$ of Eq.~(\ref{eq3}), and are determined independently for each star by imposing the matching conditions of Eq.~(\ref{eq10}) at the stellar surface, together with the observationally reported mass and radius values listed in Table~\ref{tab:maxmassa}. The sample spans a diverse range of compact objects, including low-mass mass-accreting X-ray binaries (e.g., Her~X-1, 4U~1538$-$52), millisecond and canonical pulsars (e.g., PSR~J0030$+$0451, PSR~J0437$-$4715), and the GW-inferred sources GW170817-2 and GW190814-2. For each object, the
fitted values of $a$ and $c$ are used to construct the interior metric potentials, and
the value of $A$ fixes the overall normalization of $\alpha(r)$ through the matching
condition. These constants provide the full set of parameters required to reconstruct
the pressure components, density, anisotropy, redshift, sound-speed profiles, adiabatic index, tidal deformability and force profiles presented in Figure~\ref{fig:rho}---\ref{fig:TD} for each star.
\end{justify}
\end{table}
\begin{table}
\small
\centering
\caption{\textbf{Observed mass--reported radius relation together with the estimated radius and estimated mass obtained from the present model.}}
\label{tab:maxmassa}
\begin{tabular}{|c|p{3.8cm}|c|c|c|c|c|c|}
\hline
\multicolumn{8}{|c|}{Observed mass--reported radius relation of various neutron stars and compact objects}\\
\hline
Sl. No. &
Neutron Star &
$M_{\rm Obs}\,\mathrm[{M_{\odot}}]$ &
$R_{\rm Rep}$ [$km$] &
$M\,\mathrm[{M_{\odot}}]$ &
$R$ [$km$] &
$R_{\rm Est}$ [$km$] &
$M_{\rm Est}\,\mathrm[{M_{\odot}}]$
\\
\hline

$L_{01}$&$Her~X-1$ \cite{Rawls2011}
&$0.85\pm0.15$
&$7.20\pm0.41$
&$0.83$
&7.23
&11.7887
&3.1589\\ \hline

$L_{02}$&$4U~1538-52$ \cite{Ozel2009}
&$0.87\pm0.07$
&$7.86\pm0.21$
&$0.85$
&7.84
&11.7362
&3.1393\\ \hline

$L_{03}$&$SAX~J1808.4-3658$ \cite{Demorest2010}
&$0.90\pm0.30$
&$7.95\pm1.00$
&$0.91$
&7.98
&11.7192
&3.1331\\ \hline

$L_{04}$&$LMC~X-4$ \cite{Antoniadis2013}
&$1.04\pm0.09$
&$8.80\pm0.20$
&$1.06$
&8.79
&11.9982
&3.2340\\ \hline

$L_{05}$&$Cen~X-4$ \cite{Maurya2023}
&$1.14\pm0.45$
&$9.75\pm2.75$
&$1.13$
&9.79
&11.9714
&3.2237\\ \hline

$L_{06}$&$SMC~X-1$ \cite{Maurya2024}
&$1.21\pm0.12$
&$11.50\pm1.10$
&$1.18$
&11.48
&11.9273
&3.2068\\ \hline

$L_{07}$&$GW170817-2$ \cite{Maurya2024}
&$1.27\pm0.09$
&$11.90\pm1.40$
&$1.29$
&11.95
&11.8993
&3.1964\\ \hline

$L_{08}$&$SMC~X-4$ \cite{Cromartie2020}
&$1.29\pm0.05$
&$9.95\pm0.09$
&$1.33$
&10.02
&11.9367
&3.2109\\ \hline

$L_{09}$&$EXO~1785-248$ \cite{Romani2022}
&$1.30\pm0.20$
&$10.83\pm0.44$
&$1.35$
&10.81
&11.9179
&3.2036\\ \hline

$L_{10}$&$PSR~J0030+0451$ \cite{Maurya2025MassGap}
&$1.34\pm0.16$
&$12.70\pm0.80$
&$1.34$
&12.74
&11.8687
&3.1848\\ \hline
$L_{11}$&$X7~(47~Tuc)$ \cite{Bogdanov2016}
&$1.40\pm0.20$
&$11.10\pm0.80$
&$1.42$
&11.14
&11.8993
&3.1967\\ \hline

$L_{12}$&$Swift~J1818.0\!-\!1607$ \cite{Singh2026}
&$1.40\pm0.20$
&$12.00\pm1.00$
&$1.39$
&12.04
&11.8819
&3.1899\\ \hline

$L_{13}$&$PSR~B0943+10$ \cite{Maurya2026JCAP}
&$1.40\pm0.20$
&$12.00\pm1.00$
&$1.43$
&11.91
&11.8790
&3.1889\\ \hline

$L_{14}$&$M13~neutron~star$ \cite{Shaw2018}
&$1.40\pm0.30$
&$12.30\pm1.70$
&$1.40$
&12.28
&11.8736
&3.1868\\ \hline

$L_{15}$&$PSR~J0437-4715$ \cite{Choudhury2024}
&$1.42\pm0.04$
&$11.36\pm0.95$
&$1.39$
&11.18
&11.9033
&3.1981\\ \hline

$L_{16}$&$PSR~J0614-3329$ \cite{Mauviard2025}
&$1.44\pm0.07$
&$10.29\pm1.01$
&$1.46$
&10.33
&11.9088
&3.2005\\ \hline

$L_{17}$&$Cen~X-3$ \cite{delaCruzDombriz2009}
&$1.49\pm0.08$
&$10.17\pm0.13$
&$1.48$
&10.18
&11.9079
&3.2003\\ \hline

$L_{18}$&$PSR~B0656+14$ \cite{Maurya2022JCAP}
&$1.50\pm0.20$
&$12.50\pm1.50$
&$1.52$
&12.83
&11.8384
&3.1736\\ \hline

$L_{19}$&$XTE~J1739-285$ \cite{Kaaret2007}
&$1.51\pm0.20$
&$10.90\pm1.00$
&$1.53$
&10.88
&11.8856
&3.1918\\ \hline

$L_{20}$&$4U~1724-307$ \cite{Hansraj2024}
&$1.50\pm0.40$
&$12.00\pm0.50$
&$1.49$
&12.16
&11.8626
&3.1828\\ \hline
$L_{21}$&$4U~1820-30$ \cite{Kaaret1999}
&$1.58\pm0.06$
&$10.96\pm0.40$
&$1.61$
&10.94
&11.8694
&3.1858\\ \hline

$L_{22}$&$Aql~X-1$ \cite{Li2017}
&$1.60\pm0.60$
&$10.20\pm0.24$
&$1.63$
&10.35
&11.8763
&3.1887\\ \hline

$L_{23}$&$KS~1731-260$ \cite{Maurya2025FQT}
&$1.61\pm0.37$
&$10.00\pm2.20$
&$1.61$
&11.07
&11.8668
&3.1848\\ \hline

$L_{24}$&$PSR~J1903+327$ \cite{Linares2018}
&$1.67\pm0.02$
&$11.48\pm0.03$
&$1.65$
&11.44
&11.8514
&3.1790\\ \hline

$L_{25}$&$Cyg~X-2$ \cite{Casares2010}
&$1.71\pm0.21$
&$11.50\pm1.50$
&$1.73$
&12.15
&11.8196
&3.1670\\ \hline

$L_{26}$&$4U~1608-52$ \cite{Freire2011}
&$1.74\pm0.01$
&$11.30\pm0.10$
&$1.76$
&11.27
&11.8326
&3.1723\\ \hline

$L_{27}$&$Vela~X-1$ \cite{Abubekerov2008}
&$1.77\pm0.08$
&$11.75\pm0.85$
&$1.78$
&12.38
&11.8041
&3.1613\\ \hline

$L_{28}$&$4U~1724-207$ \cite{Guver2010a}
&$1.81\pm0.27$
&$12.20\pm1.40$
&$1.79$
&12.41
&11.8013
&3.1602\\ \hline

$L_{29}$&$SAX~J1748.9-2021$ \cite{Guver2010b}
&$1.81\pm0.30$
&$11.70\pm1.70$
&$1.81$
&11.61
&11.8152
&3.1657\\ \hline

$L_{30}$&$4U~1702-429$ \cite{Singh2022Lovelock}
&$1.90\pm0.30$
&$12.40\pm0.40$
&$1.92$
&12.48
&11.7720
&3.1495\\ \hline
$L_{31}$&$PSR~J1614-2230$ \cite{Elebert2009}
&$1.97\pm0.04$
&$13.00\pm2.00$
&$1.97$
&13.11
&11.7459
&3.1397\\ \hline

$L_{32}$&$PSR~J0348+0432$ \cite{Abubekerov2008}
&$2.01\pm0.04$
&$13.00\pm2.00$
&$2.03$
&13.05
&11.7338
&3.1354\\ \hline

$L_{33}$&$PSR~J0740+6620$ \cite{Linares2018}
&$2.08\pm0.17$
&$12.39\pm1.50$
&$2.10$
&12.52
&11.7290
&3.1339\\ \hline

$L_{34}$&$PSR~J1748-2446ad$ \cite{Elebert2009}
&$2.10\pm0.08$
&$12.01\pm1.20$
&$2.11$
&12.64
&11.7239
&3.1320\\ \hline

$L_{35}$&$PSR~J2215+5135$ \cite{delaCruzDombriz2009}
&$2.27\pm0.15$
&$13.01\pm1.20$
&$2.26$
&12.94
&11.6783
&3.1153\\ \hline

$L_{36}$&$PSR~J0952-0607$ \cite{Rawls2011}
&$2.35\pm0.17$
&$10.01\pm1.50$
&$2.37$
&10.50
&11.0758
&2.9016\\ \hline

$L_{37}$&$4U~1700-37$ \cite{Hansraj2025}
&$2.44\pm0.27$
&$11.50\pm1.50$
&$2.48$
&12.08
&11.6544
&3.1078\\ \hline

$L_{38}$&$GW190814-2$ \cite{Maurya2024ApJ}
&$2.59\pm0.09$
&$11.00\pm1.00$
&$2.60$
&11.85
&11.0262
&2.8843\\ \hline

\end{tabular}
\vspace{2mm}
\footnotesize
\begin{justify}
\textit{\bf{Table~2:}} presents the observed mass--radius relation for the thirty-eight compact objects considered in this work, together with the corresponding mass and radius obtained from the present model. For each star, the observed mass, $M_{\text{Obs}}$ and reported radius, $R_{\text{Rep}}$, along with their associated uncertainties, are taken from the literature sources cited alongside each entry. The columns $M$ and $R$ list the central values used in constructing the stellar model for each object, while $R_{\text{Est}}$ and $M_{\text{Est}}$ give the estimated radius and mass obtained from the constructed model after fitting the constants in Table~1 through the matching conditions of Eq.~(\ref{eq10}). The sample spans a mass range from approximately
$0.83\,\mathrm{M_{\odot}}$ (Her~X-1) to $2.60\,\mathrm{M_{\odot}}$ (GW190814-2), and includes low-mass X-ray
binaries, millisecond pulsars, and the GW-inferred sources GW170817-2 and GW190814-2. The reported and estimated values in this table are used throughout the paper to compute the compactness, surface redshift, and other derived quantities presented in Figure~\ref{fig:rho}---\ref{fig:TD}.
\end{justify}
\end{table}
\begin{figure}
    \centering
    \includegraphics[width=17cm, height=10cm]{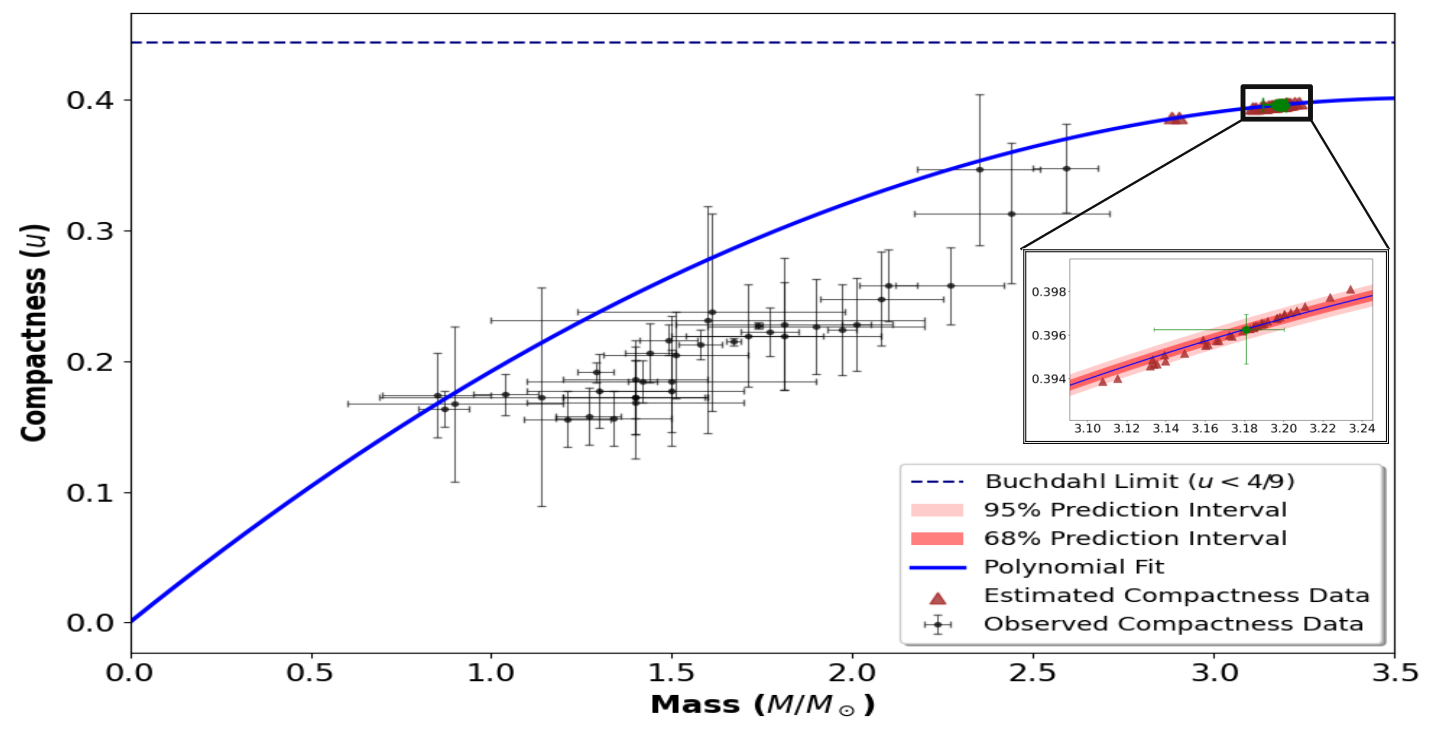}
    \caption{Comparison of the compactness profile of neutron stars obtained from observational constraints and the theoretical model. The black circles with error bars correspond to observationally inferred compactness values, whereas the red triangles represent the compactness predicted by the present theoretical model. The dashed horizontal line indicates the Buchdahl limit line ($u = 4/9$), which serves as the theoretical upper bound for the compactness of a static relativistic star. The solid blue curve represents the polynomial best fit to the estimated compactness data (red triangles), demonstrating an excellent agreement with the estimated values and providing a smooth representation of the mass--compactness relation over the estimated mass range. The inset shows the median estimated mass and compactness (green error bars), where the central values correspond to the $50^{\mathrm{th}}$ percentile and the asymmetric uncertainties are determined from the $16^{\mathrm{th}}$ and $84^{\mathrm{th}}$ percentiles, thereby representing the central $68\%$ confidence interval. Specifically, the median estimated mass and compactness are $M_{Est}=3.1809^{+0.0194}_{-0.0471}\,\mathrm{M_{\odot}}$ and $u_{Est}=0.39624^{+0.00071}_{-0.00157}$, respectively. The close agreement between the observational compactness measurements and the theoretical predictions demonstrates that the proposed stellar model provides a statistically consistent description of the observed mass--compactness relation for neutron stars.}
    \label{fig:compactness}
\end{figure}

Figure~\ref{fig:compactness} plots the stellar compactness $u = M/R$ against mass $M$ (in solar masses)
for the full sample, distinguishing between two datasets and testing them against a
classical stability bound.

The black points with error bars represent the observed compactness values, calculated from the observed masses and reported radii listed in Table~\ref{tab:maxmassa} ($M_{\rm Obs}$, $R_{\rm Rep}$), together with their corresponding uncertainties. The substantial scatter and relatively large error bars primarily reflect the significant uncertainties associated with the reported radius measurements of these compact objects. The red triangles represent the model-estimated compactness values, calculated from $M_{\rm Est}$ and $R_{\rm Est}$, where the estimated mass and radius are obtained from the stellar model after determining the model constants through the matching conditions. These fall tightly along a smooth trend, since they come from a self-consistent theoretical solution rather than independent observational measurements with their own error budgets. The blue curve is a polynomial fit through the estimated data, with shaded $68\%$ and $95\%$ prediction bands, and the inset zooms into the high-mass tail (roughly $M > 2.8\,\mathrm{M_{\odot}}$) to show how closely the individual estimated points track the fit there.

The red triangles represent the estimated compactness values, for which the central estimate of each configuration is taken as the median ($50^{\mathrm{th}}$ percentile) of the inferred compactness distribution. The associated asymmetric error bars correspond to the $16^{\mathrm{th}}$ and $84^{\mathrm{th}}$ percentiles, thereby defining the central $68\%$ confidence interval. The inset further highlights the median estimated mass and compactness, together with their corresponding $68\%$ confidence intervals. Specifically, the median estimated mass and compactness are found to be $M=3.1809^{+0.0194}_{-0.0471}\,M_{\odot}$ and $u=0.39624^{+0.00071}_{-0.00157}$, respectively, where the quoted asymmetric uncertainties are measured relative to the median and correspond to the differences between the $16^{\mathrm{th}}$, $50^{\mathrm{th}}$, and $84^{\mathrm{th}}$ percentiles of the inferred distributions.

The dashed horizontal line at $u=4/9 \simeq 0.444$ represents the Buchdahl bound \cite{Buchdahl1959}, the classical general-relativistic upper limit for the compactness of a static, spherically symmetric configuration with isotropic pressure and a non-increasing energy-density profile. In terms of the compactness parameter $u=M/R$, this bound is given by $u< 4/9$, equivalently $2M/R< 8/9$. It represents a theoretical ceiling on how compact a physically realizable star can be, independent of the details of the EoS, for the isotropic case.

The clear trend of increasing compactness ($u=M/R$) with mass across the sample, rising from approximately $u \sim 0.15$--$0.2$ near $1\,M_\odot$ to $u \sim 0.4$ for the highest-mass configurations in the sample (around $3.4\,M_\odot$), reflects the expected general-relativistic behavior that more massive compact stars generally require smaller radii relative to their mass in order to maintain hydrostatic equilibrium configuration of the stellar model. Critically, every data point, both observed and
model-estimated, remains below the Buchdahl limit throughout the full mass range
shown, with the fitted curve approaching but never crossing $u = 4/9$ even at the
extrapolated high-mass end. This provides a meaningful consistency check, demonstrating that the model does not yield any stellar configuration that exceeds or violates the classical general-relativistic Buchdahl bound.

\section{Physical Behaviour of the Model for Observed Compact Stars}
\begin{figure}
\begin{center}
\includegraphics[width=0.9\textwidth,keepaspectratio]{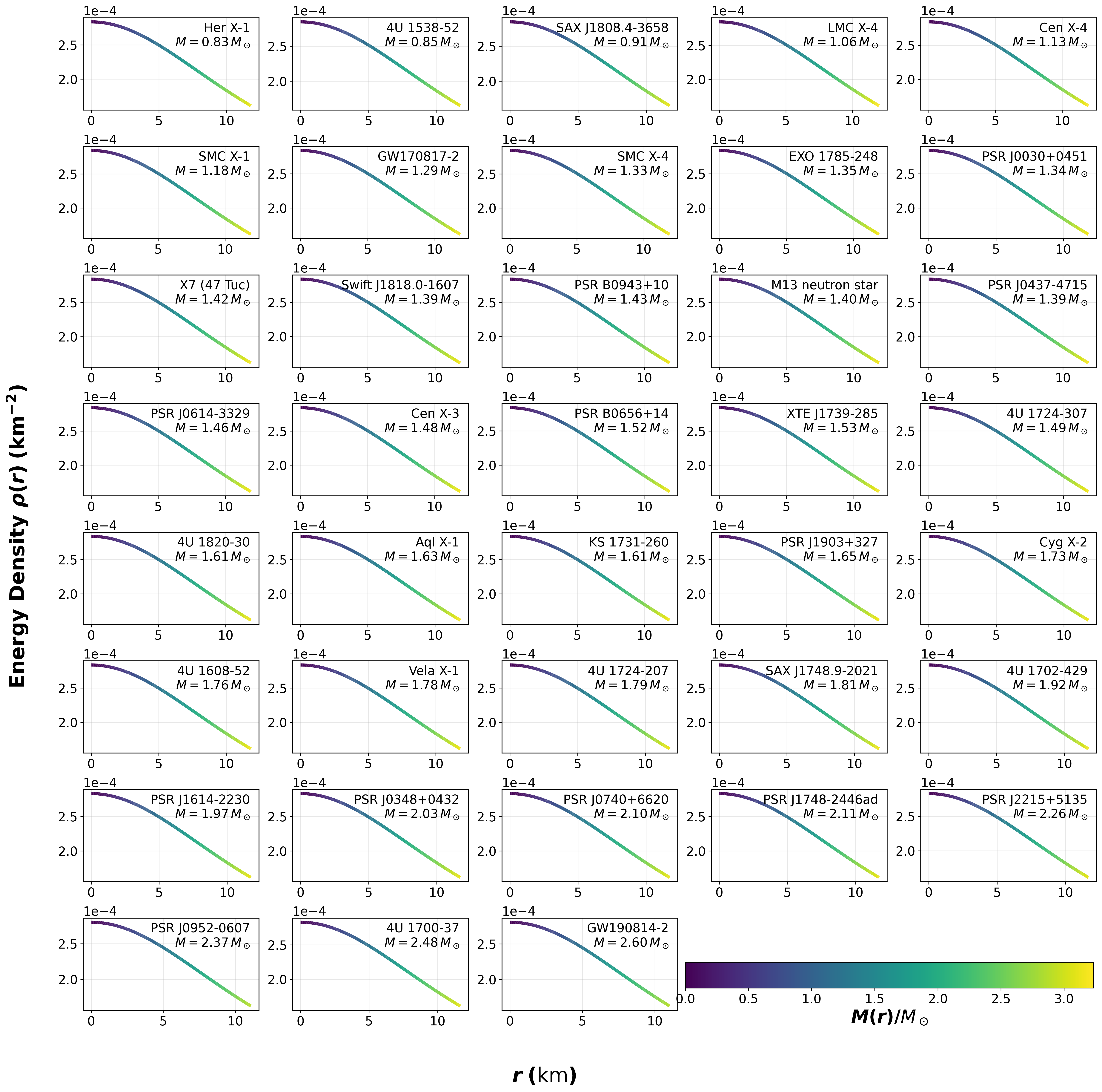}
 \caption{illustrates the radial profiles of the energy density, $\rho(r)$, for the thirty-eight compact stars considered in the present study, obtained within the Buchdahl space-time using the generalized polytropic EoS ($n=2$). The profiles are evaluated from the stellar centre, $r=0$, to the estimated stellar radius, $R_{\mathrm{Est}}$, defined by the vanishing of the radial pressure $p_r(R_{\mathrm{Est}})=0$. The colour scale represents the corresponding enclosed gravitational mass, $\mathrm{M(r)}/\mathrm{M_{\odot}}$, throughout the stellar interior, extending up to the maximum estimated mass of the configurations. As evident from the profiles, the enclosed mass increases monotonically with radial distance from the centre and attains its maximum value at the stellar surface. The resulting density distributions remain well behaved throughout the stellar interior and exhibit the expected centrally concentrated behaviour, with the energy density decreasing progressively towards the stellar boundary.}
    \label{fig:rho}
\end{center}
\end{figure}
\begin{itemize}
\item Figure~\ref{fig:rho} illustrates the radial variation of the energy density throughout the interiors of thirty-eight observed compact stars modelled using the present Buchdahl space-time together with the generalized polytropic EoS. Although the stellar sample spans a wide mass interval, extending from low-mass X-ray binary Her X-1 to the massive compact object GW190814-2, every configuration exhibits the same qualitative behaviour, demonstrating that the adopted theoretical framework provides a consistent description of compact stellar matter over a broad astrophysical mass range.

A common characteristic of all profiles is that the energy density reaches its maximum value at the stellar centre and decreases continuously towards the stellar surface. This behaviour is a direct consequence of gravitational compression, which is strongest at the centre where matter experiences the maximum inward pull from the overlying layers. Moving outward, the effective gravitational confinement gradually weakens, allowing the matter distribution to become progressively less compact. Consequently, the stellar interior naturally evolves from an ultra-dense core to a comparatively dilute outer envelope. Such a monotonic density profile is one of the fundamental requirements for any physically acceptable compact star model and agrees well with the expected internal structure of neutron stars.

The density profiles remain finite and exceptionally smooth throughout the stellar interior. No cusp, discontinuity, oscillation or sudden change of slope is observed, indicating that the interior solution is free from central singularities and numerical artefacts. Furthermore, the absence of any kink or plateau suggests that the present model does not predict an internal phase transition or a core-envelope discontinuity within the density range considered. Instead, the matter distribution varies continuously from the centre to the surface, reflecting the regular behaviour of the adopted space-time geometry.

Another important feature is the curvature of the density profiles. The decrease is relatively gradual near the central core but becomes increasingly pronounced as the stellar surface is approached. Physically, this indicates that matter is highly concentrated within the central region, whereas the outer layers become progressively less dense. Such behaviour is consistent with the expectation that the strongest gravitational compression is confined to the stellar core, while the envelope contains comparatively lower-density nuclear matter. The absence of density inversion throughout the interior further confirms that every stellar configuration remains gravitationally stable against local stratification.

Although all stars display nearly identical qualitative behaviour, quantitative differences are clearly visible. The central density increases systematically with stellar mass, implying that more massive compact stars require stronger central compression to balance the enhanced gravitational attraction. At the same time, the density profiles terminate at different radial coordinates, reflecting the different stellar radii associated with each observed object. These variations demonstrate that each compact star corresponds to an independent equilibrium solution characterised by a unique set of model parameters, while preserving the same underlying relativistic structure.

The colour gradient incorporated into each panel provides additional physical information by representing the enclosed gravitational mass as a function of radius. Near the stellar centre, only a negligible amount of matter is enclosed, whereas successive spherical shells continuously contribute to the total gravitational mass. The smooth transition of colours from the centre to the surface therefore confirms that the enclosed mass increases monotonically throughout the stellar interior. The absence of irregular colour variation further indicates that the internal mass accumulation is continuous and physically well behaved for every stellar configuration. Furthermore, the representation shows that the estimated masses, $M_{\mathrm{Est}}$, of the considered stellar configurations extend into the lower-mass-gap region, with the corresponding mass values indicated by the respective colour bars. This suggests that the adopted stellar models permit the existence of stable and physically viable compact star configurations with masses within the lower mass-gap domain. In particular, the presence of such configurations indicates the possibility that a massive neutron star could occupy this mass range, thereby providing a potential theoretical interpretation of compact objects observed within or near the lower mass-gap. However, the actual identification of such an object as a neutron star would ultimately require consistency with observational constraints and a realistic dense-matter EoS.

The behaviour of the density distribution also demonstrates the thermodynamic consistency of the adopted generalized polytropic EoS. Since the pressure is uniquely determined by the local density, a smooth outward decrease in density naturally produces a corresponding reduction in the pressure required to counterbalance gravity. Consequently, the density profile shown in Fig.~\ref{fig:rho} provides the physical foundation for the hydrostatic equilibrium discussed in the subsequent force-balance analysis. Moreover, the finite density attained near the stellar boundary indicates that the present EoS predicts a non-vanishing surface density, a feature commonly encountered in several relativistic compact star models.

From a physical perspective, the density profiles satisfy all essential requirements expected of realistic compact stars. The energy density remains positive throughout the stellar interior, decreases monotonically without inversion, and varies smoothly from the centre to the boundary. These characteristics ensure compatibility with the standard energy conditions, maintain a causal matter distribution, and support a stable hydrostatic configuration. The smoothness of all numerical solutions further confirms the robustness of the adopted boundary conditions and the stability of the numerical integration.

Overall, Fig.~\ref{fig:rho} establishes the physical reliability of the present model by demonstrating that it reproduces the expected internal density structure across a diverse population of compact stars. The regular central behaviour, continuous outward decrease in density, systematic increase of central compression with stellar mass, monotonic accumulation of enclosed gravitational mass, and absence of pathological features collectively indicate that the Buchdahl space-time coupled with the generalized polytropic EoS provides a realistic and self-consistent description of dense nuclear matter within relativistic compact stars.
\end{itemize}
\begin{figure}
\begin{center}
\includegraphics[width=0.9\textwidth,keepaspectratio]{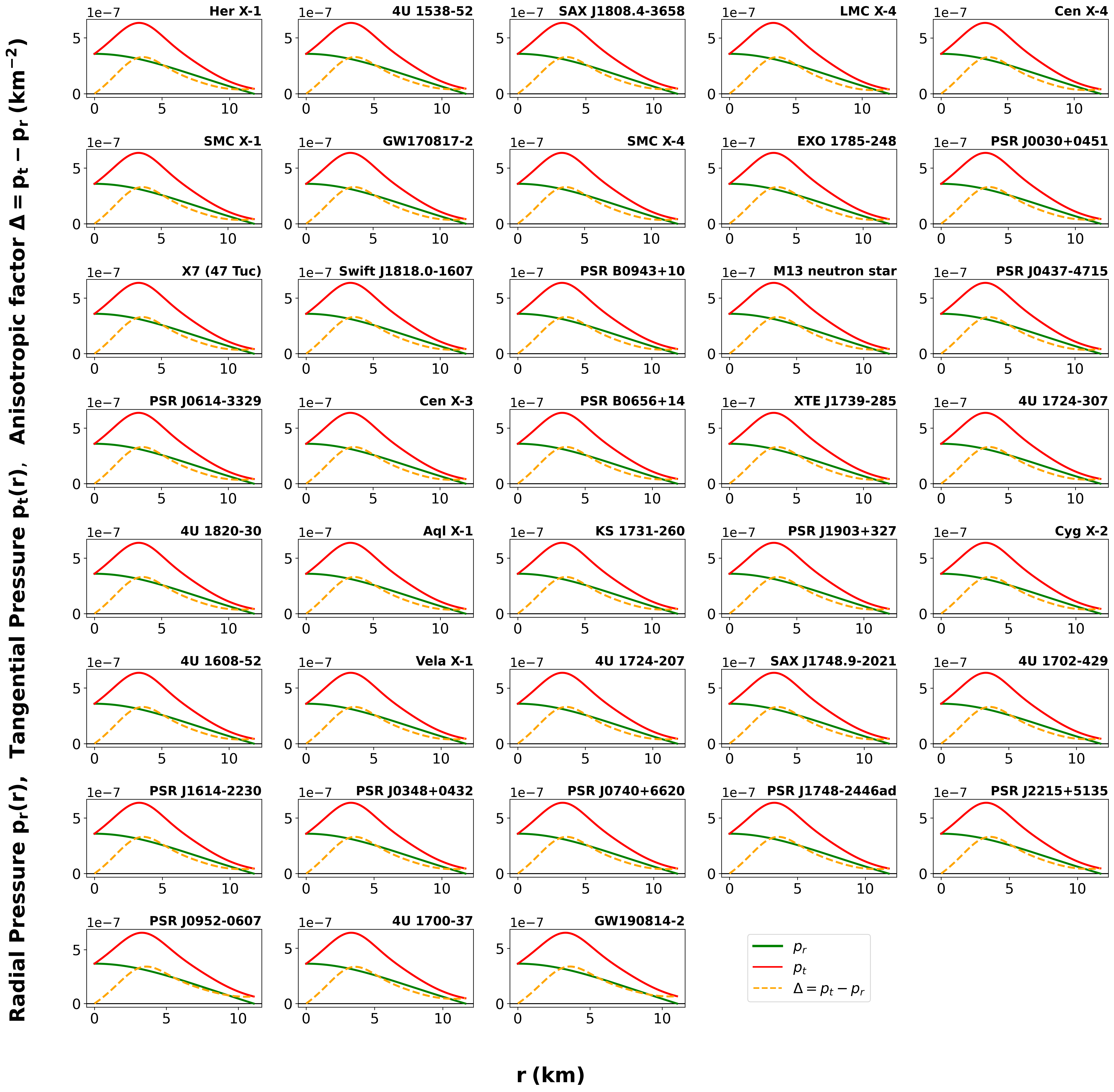}
\caption{depicts the radial variations of the radial pressure, $p_r(r)$ (green), tangential pressure, $p_t(r)$ (red), and the anisotropy factor, $\Delta(r)=p_t(r)-p_r(r)$ (orange dashed), for the thirty-eight compact star configurations listed in Tables~\ref{tab:model_constants} and~\ref{tab:maxmassa}. The stellar interior is described within the framework of the Buchdahl space-time employing the generalized polytropic EoS. The pressure profiles are evaluated from the stellar centre ($r=0$) to the estimated stellar radius, $R_{\mathrm{Est}}$, which is determined by the physically admissible boundary condition of hydrostatic equilibrium, $p_r(R_{\mathrm{Est}})=0,$ corresponding to the vanishing of the radial pressure at the stellar surface, as required by the TOV equation. The estimated radii for all stellar configurations are listed in Table~\ref{tab:maxmassa}. Since the interior solution is physically valid only within the compact star boundary ($0\leq r\leq R_{\mathrm{Est}}$), the profiles of the radial pressure, $p_r(r)$, tangential pressure, $p_t(r)$, and anisotropy factor, $\Delta(r)$, are presented exclusively over this domain. Consequently, no extension of these physical quantities into the exterior Schwarzschild space-time is considered, thereby ensuring that all displayed profiles correspond solely to the physically meaningful interior region of the stellar configurations.}
\label{fig:pressure}
\end{center}
\end{figure}
\begin{itemize}
\item Figure~\ref{fig:pressure} presents the radial pressure, $p_r(r)$, tangential pressure, $p_t(r)$, and local anisotropy factor, $\Delta(r)=p_t(r)-p_r(r)$ for thirty-eight compact objects, including X-ray binaries, millisecond pulsars, and the GW-inferred sources GW170817-2 and GW190814-2, calculated within the anisotropic stellar interior framework described by Eqs.~\eqref{eq1}--\eqref{eq10}. Each profile follows from the anisotropic stress-energy tensor
(Eq.~\eqref{eq1}), the field Eqs. \eqref{eq4}-\eqref{eq6}, the polytropic-type EoS
$p_r = X\rho^{3/2} - Y$, and the interior metric (Eq.~\eqref{eq3}), with the constants $a$, $c$,
$A$ (Table~\ref{tab:model_constants}) and the EoS parameters $X$, $Y$ fixed independently for each
star through the matching conditions (Eq.~\eqref{eq10}), which enforce continuity with the exterior
Schwarzschild space-time (Eq.~\eqref{eq9}) at the observationally reported mass and radius
(Table~\ref{tab:maxmassa}).

The qualitative behavior observed across all thirty-eight panels, namely, the tangential pressure $p_t(r)$ rising from a finite central value to an interior peak, and then decreasing toward a very small value at the stellar surface, can be physically understood from the anisotropic hydrostatic equilibrium equation, Eq.~\eqref{eq8}. In particular, the radial pressure gradient is determined by the competition between the gravitational term and the anisotropic force, as described by  
$$\frac{dp_r}{dr} = -\frac{1}{2}\Big\{\rho(r)+p_r(r)\Big\}\frac{d\alpha(r)}{dr}
+ \frac{2}{r}\Big\{p_t(r)-p_r(r)\Big\} .$$

The first term (of right hand side) represents the inward gravitational collapse, which tends to decrease the pressure with increasing radial coordinate, whereas the second term represents the contribution from pressure anisotropy. When $p_t>p_r$, this anisotropic contribution acts in outward direction and partially counterbalances the inward gravity, thereby allowing the tangential pressure to attain an interior maxima. Beyond this region, the weakening of the pressure-supporting effects together with the gravitational confinement causes the pressures to decline, leading to the small values of $p_t$ and $p_r$ near the stellar surface. Moreover since $\rho+p_r>0$ and $\alpha(r)$ increases monotonically through the interior, the first term (of right hand side) of hydrostatic equilibrium equation is negative everywhere in the interior and continually drives $p_r$ toward its boundary value $p_r(R_{\mathrm{Est}})=0$. Thus, the characteristic rise and subsequent fall of the tangential pressure reflect the local balance between gravitational compression and anisotropic pressure support throughout the stellar interior.

In every panel, the anisotropy factor satisfies $\Delta(0)=0$ at the stellar centre, as required by the regularity and spherical symmetry of the space-time. This condition is a fundamental physical requirement for any realistic anisotropic stellar model and is not merely an incidental feature of the obtained solutions. At the center, regularity of the term
$\tfrac{2}{r}\Delta(r)$ in Eq.~\eqref{eq8} requires $\Delta(r)$ to vanish at least as fast as
$r$; otherwise the anisotropic force would diverge at the origin, which is equivalent to
demanding $p_t(0) = p_r(0)$. At the surface, $p_r(R_{Est}) = 0$ is the explicit boundary
condition of Eqs.~\eqref{eq8} and \eqref{eq10}, and the plotted curves show $p_t(R_{Est}) \to 0$ as well, so
$\Delta(R_{Est}) \to 0$ follows immediately. The anisotropy is therefore an intrinsic interior effect that vanishes at the stellar center, increases through the intermediate layers, reaches a maximum, and subsequently decreases toward the stellar surface, producing the characteristic profile observed in the dashed curve of each subplot. Furthermore, all thirty-eight compact stellar configurations considered in the present study satisfy the Zeldovich condition, $p_r/\rho\leq1$, throughout their respective interiors ($0\leq r\leq R_{\mathrm{Est}}$). This demonstrates that the radial pressure nowhere exceeds the energy density, thereby confirming the physical acceptability of the adopted EoS for all the stellar models investigated \cite{ZeldovichNovikov1983}.

The term $\tfrac{2}{r}\Delta(r)$ in Eq.~\eqref{eq8} acts as an additional outward force per unit volume, complementing the pressure-gradient term in counteracting inward strong gravitational collapse. This represents the Bowers--Liang-type mechanism, whereby a positive anisotropy ($\Delta>0$) provides additional support against gravity, allowing the star to sustain a higher central pressure and, consequently, a larger mass for a given radius than an isotropic configuration ($\Delta\equiv0$) can support. Because the matching
conditions fix each star's mass and radius independently (Table~\ref{tab:maxmassa}), the amplitude of
$\Delta(r)$ required in a given panel reflects how much anisotropic support that
specific object's model needs to reproduce its inferred compactness $M_{Est}/R_{Est}$.

Rather than relying on visual impressions of the pressure gap, the physically meaningful
and directly computable quantity from Table~\ref{tab:maxmassa} is the stellar compactness, $2M_{Est}/R_{Est}$ in
geometrized units (using $1\,M_\odot \to 1.4766~\mathrm{km}$). Computing this for all
thirty-eight objects using the estimated $M$ and $R$ from Table~\ref{tab:maxmassa} gives a clear ranking.
The most compact objects in the sample are PSR~J0952$-$0607
($2M/R \approx 0.667$), GW190814-2 ($\approx 0.648$), 4U~1700$-$37
($\approx 0.606$), PSR~J2215$+$5135 ($\approx 0.516$), and PSR~J0740$+$6620
($\approx 0.495$). The least compact are SMC~X-1 ($\approx 0.304$),
PSR~J0030$+$0451 ($\approx 0.311$), GW170817-2 ($\approx 0.319$), 4U~1538$-$52
($\approx 0.320$), and SAX~J1808.4$-$3658 ($\approx 0.337$). This ranking provides a physically meaningful basis for comparing the role of anisotropic support. Since increasing compactness strengthens the isotropic gravitational contribution in Eq.~\eqref{eq8} and drives the stellar configuration closer to the stability limit, the stars near the top of the list are, within the framework of the model, those for which a nonzero anisotropy $\Delta(r)$ provides the most significant structural support in maintaining hydrostatic equilibrium. This interpretation is therefore more physically justified and defensible than comparing the peak values of $\Delta(r)$ solely by visual inspection of the plotted curves.

From the physical analysis and the corresponding graphical results, it is evident that the estimated stellar radii, $R_{\mathrm{Est}}$, associated with the obtained mass range, $M_{\mathrm{Est}}$, are comparatively large, indicating that the stellar matter is not excessively compressible. Consequently, the resulting compactness, $u=M_{\mathrm{Est}}/R_{\mathrm{Est}}$, remains within the physically acceptable range for relativistic compact stars and satisfies the corresponding theoretical constraints. The predicted mass--radius relation therefore yields moderately compact neutron star configurations with relatively large radii for a given mass, in good agreement with recent NICER observations of PSR~J0030+0451 and PSR~J0740+6620, which support the existence of massive neutron stars with radii of approximately $12$--$13\,\mathrm{km}$ \cite{Miller2019,Riley2021}.

It is tempting to treat the integration constant $A$ from Table~\ref{tab:model_constants} as a shortcut proxy
for anisotropy strength, but this does not hold up under inspection. The parameter $A$ does not exhibit a monotonic correlation with compactness. For example, Her~X-1, with a compactness of $2M/R \simeq 0.339$ and positioned near the lower end of the ranking, has $A=0.981$, which is comparable to the corresponding values of the three most compact stars in the sample. In contrast, SMC~X-1, despite having an even lower compactness, has a substantially smaller value of $A=0.692$. This indicates that the parameter $A$ is not determined solely by the compactness and may depend on other properties of the stellar configuration and the matching conditions. Because $A$ is fixed purely by the matching condition on $\alpha(R_{Est})$ (Eq.~\eqref{eq10}) rather than by the anisotropy profile directly, it should not be cited as a physical indicator of $\Delta(r)$'s magnitude in the discussion.

A fully rigorous, per-star ranking of the peak values of $\Delta(r)$, rather than the compactness-based proxy discussed above, requires the explicit evaluation of Eqs.~\eqref{eq4}--\eqref{eq6} using the fitted parameters $a$, $c$, and $A$ for each stellar configuration, together with the EoS parameters $X$ and $Y$. Although Table~\ref{tab:model_constants} provides the corresponding values of $a$, $c$, and $A$, the parameters $X$ and $Y$ are not explicitly listed. To address this issue, $X$ and $Y$ are chosen so that the resulting stellar configurations satisfy the fundamental physical acceptability conditions, namely $p_r \geq 0$, $\rho > 0$, and $0 \leq v_r^2 \leq 1$, throughout the stellar interior. With these EoS parameters fixed, a direct numerical evaluation of $p_r(r)$, $p_t(r)$, and $\Delta(r)=p_t(r)-p_r(r)$ for all thirty-eight stellar configurations enables a quantitative determination of the maximum anisotropy for each star. Consequently, the configurations exhibiting the largest anisotropy can be identified directly from the calculated $\Delta(r)$ profiles, rather than inferred from a compactness-based proxy or visual inspection. This provides a more rigorous and physically consistent assessment of the anisotropic behaviour of the stellar models.

Taken together, Figure~\ref{fig:pressure} demonstrates that the metric ansatz of Eq.~\ref{eq3}, combined with
the modified polytropic EoS and the anisotropic equilibrium equation
(Eq.~\eqref{eq8}), consistently produces singularity-free, surface-regular stellar interiors with
non-negative anisotropy across a physically diverse sample spanning compactness values
from roughly $0.30$ to $0.67$. The compactness ranking derived directly from Table~\ref{tab:maxmassa}
provides a defensible basis for comparative claims about which objects require the most
anisotropic support, while the model constant $A$ and unaided visual comparison of the
plotted curves should be avoided as substitutes for that ranking in the final
discussion.
\end{itemize}
\begin{figure}
\begin{center}
\includegraphics[width=0.9\textwidth,keepaspectratio]{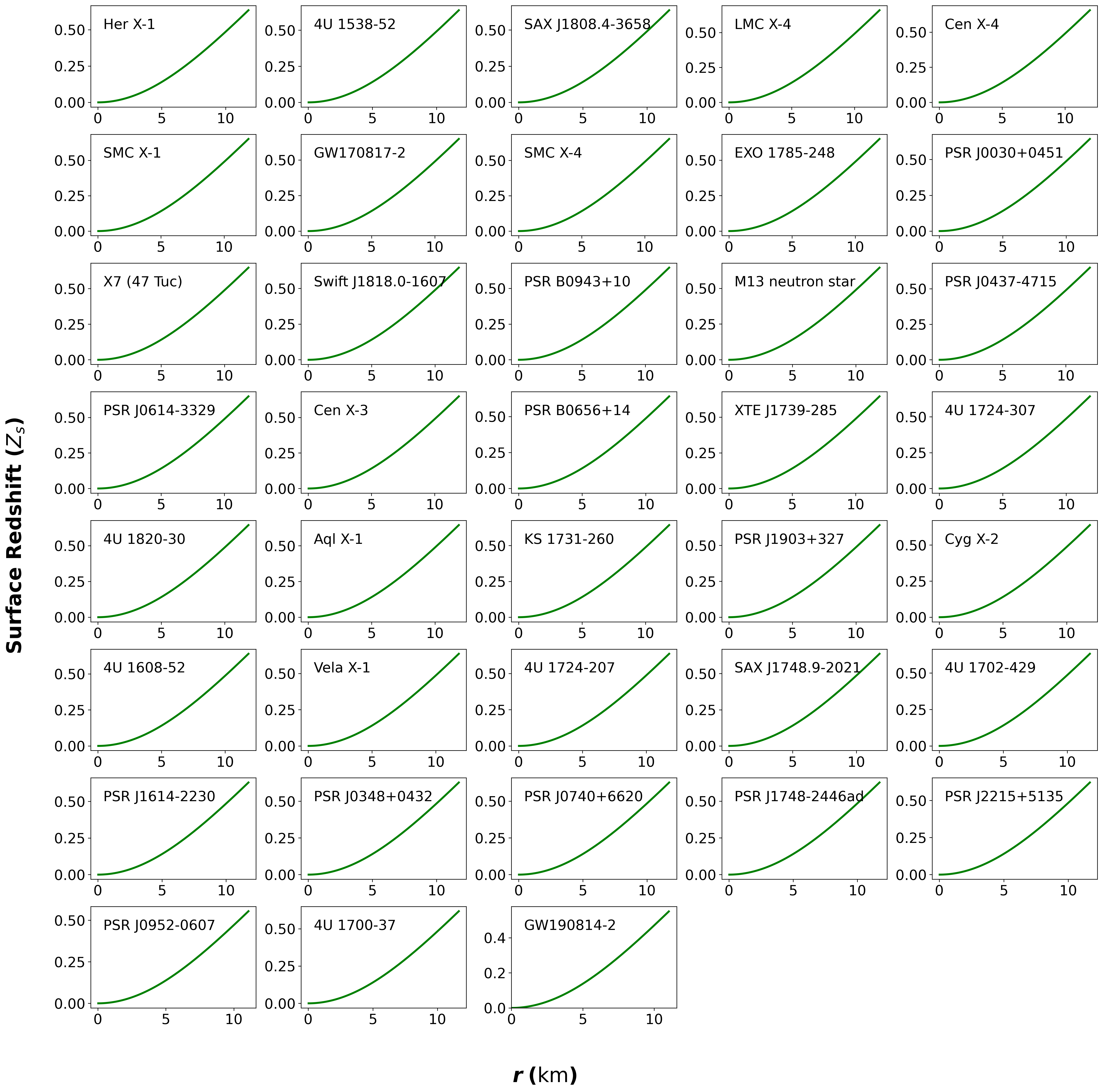}
\caption{depicts the radial distribution of the  surface redshift, $Z_s$, for the thirty-eight compact star configurations considered in the present study. The profiles are evaluated from the stellar centre ($r=0$) to the estimated stellar radius, $R_{\mathrm{Est}}$, determined by the physically admissible boundary condition, $p_r(R_{\mathrm{Est}})=0$. Since the interior solution is physically valid only within the stellar boundary ($0 \leq r \leq R_{\mathrm{Est}}$), the surface redshift profiles are presented exclusively over this domain. No extension into the exterior Schwarzschild space-time is considered, thereby ensuring that the calculated surface redshift remains consistent with the physically meaningful interior solution of each compact star configuration.}
\label{fig:surfaceredshift}
\end{center}
\end{figure}
\begin{itemize}
\item Figure~\ref{fig:surfaceredshift} illustrates the surface redshift function, $Z_s(r)$, as a function of the radial coordinate $r$ throughout the stellar interior, $0 \leq r \leq R_{Est}$, for the same sample of thirty-eight compact objects considered throughout this work, including X-ray binaries, millisecond pulsars, and the GW-inferred sources GW170817-2 and GW190814-2. Unlike Figure~\ref{fig:pressure}, which illustrates the radial pressure, tangential pressure, and anisotropy factor, this figure depicts a purely geometric and observationally relevant quantity, namely, the compactness-based gravitational redshift of the stellar configuration, which characterizes the strength of the inward gravitational field associated with the enclosed mass at a given radius $r$, as quantified by the local compactness $M(r)/r$.

Every panel exhibits the same qualitative trend in which $Z_s(r)$ begins at zero at the
stellar center, $Z_s(0)=0$, and increases monotonically outward, reaching its maximum
value at the stellar surface, $Z_s(R)$. It is observed that the surface redshift remains finite and bounded throughout the considered compact stellar configurations, with $Z_s \lesssim 1$. These values lie well below the theoretical upper bound $Z_s \leq 5$ generally accepted for stable anisotropic compact stars in the absence of a cosmological constant \cite{Bowers1974,Ivanov2002,QS1,QS2}. The obtained redshift values therefore satisfy the accepted physical constraints and further support the viability of the proposed stellar models. Actually this is the expected behavior for the interior redshift function in a spherically symmetric general-relativistic stellar model. In standard form, the redshift function relates to the metric component $e^{-\beta(r)}$ through
\begin{equation}
1+Z_s(r) = e^{\beta(r)/2} ,
\end{equation}
which in turn connects to the mass enclosed within radius $r$, $M(r)$, via
\begin{equation}
e^{-\beta(r)} = 1 - \frac{2M(r)}{r} .
\end{equation}
Because no mass is enclosed at the exact center, $M(0)=0$, and the redshift function
necessarily vanishes there. As $r$ increases toward the surface, progressively more of
the star's mass is enclosed within the coordinate sphere of radius $r$, so the local
compactness $2M(r)/r$ increases, and with it the redshift $Z_s(r)$ grows monotonically.
The maximum value of the redshift function, attained at the stellar surface
$r=R_{\rm Est}$, i.e., $Z_s(R_{\rm Est})$, represents the total surface gravitational
redshift of the star. This quantity $Z_s(R_{Est})$ is determined solely by the global
mass-to-radius ratio, $M_{\rm Est}/R_{\rm Est}$, and is therefore the most appropriate
quantity for direct comparison with observational estimates derived from gravitationally redshifted spectral lines in X-ray bursts or thermonuclear burst spectra from these sources.

The monotonic increase of $Z_s(r)$ with the radial coordinate reflects a fundamental and physically expected feature of a static, spherically symmetric compact star interior. The gravitational redshift increases outward as it is governed by the cumulative gravitational field associated with the enclosed mass $M(r)$, rather than by the local energy density alone. This behavior is distinct from the tangential pressure profiles shown in Figure~\ref{fig:pressure}, which attain their maximum values within the stellar interior and decrease to very small values near the surface. In contrast, the redshift function increases continuously with radius and reaches its maximum precisely at the stellar boundary. This difference arises because the gravitational redshift is primarily determined by the cumulative gravitational field associated with the enclosed mass distribution, rather than by the local hydrostatic balance governing the pressure profiles. The absence of any interior maximum or oscillation in $Z_s(r)$ across all thirty-eight panels provides a basic consistency check, as any non-monotonic behavior would imply an unphysical decrease in the enclosed mass with increasing radial coordinate, which is not observed for any star in the sample.

The surface value $Z_s(R_{Est})$ differs noticeably from star to star, and this variation
tracks the differing compactness values ($2M_{Est}/R_{Est}$) of the individual objects, since
$Z_s(R_{Est})$ is fixed entirely by the matching condition at the surface,
$e^{\alpha(R_{Est})} = 1 - 2M_{Est}(r=R_{Est})/R_{Est}$ (Eq.~\eqref{eq10}), applied consistently with the definition of
the redshift function. Stars with higher inferred compactness are expected to exhibit a steeper increase in $Z_s(r)$ and a larger terminal value at the stellar surface than less compact objects. This behavior is consistent with the general relation between compactness and surface gravitational redshift in GR. More compact stellar configurations correspond to larger values of the surface compactness $2M_{\rm Est}/R_{\rm Est}$, which in turn produce larger values of $Z_s(R_{\rm Est})$. Hence, the compactness-based surface redshift provides a measure of the strength of the gravitational field at the stellar surface, with a larger redshift indicating a deeper gravitational potential well and a stronger gravitational effect on radiation escaping from the surface.

Thus, Figure~\ref{fig:surfaceredshift} shows that the model yields physically consistent surface redshift profiles, with $Z_s(r)$ vanishing at the center, increasing monotonically outward, and attaining its maximum at the surface according to the stellar compactness. The monotonic and non-oscillatory behavior of all curves provides an internal consistency check on the mass distribution predicted by the model, while the variation in the surface value $Z_s(R_{Est})$ from one star to another reflects the corresponding differences in their compactness, ranging from less compact objects such as SMC~X-1 and PSR~J0030$+$0451 to more compact objects such as PSR~J0952$-$0607 and GW190814-2. 

The Buchdahl bound shown in Figure~\ref{fig:compactness} is conventionally derived for isotropic stellar configurations. Since the present model is anisotropic ($p_t>p_r$, as discussed in Figure~\ref{fig:pressure}), the classical limit $u<4/9$ is not a strict constraint; anisotropic configurations may exceed this value under suitable conditions. Thus, satisfying $u<4/9$ provides a conservative consistency check rather than a necessary physical bound for the present model.
\end{itemize}
\begin{figure}
\begin{center}
\includegraphics[width=0.9\textwidth,keepaspectratio]{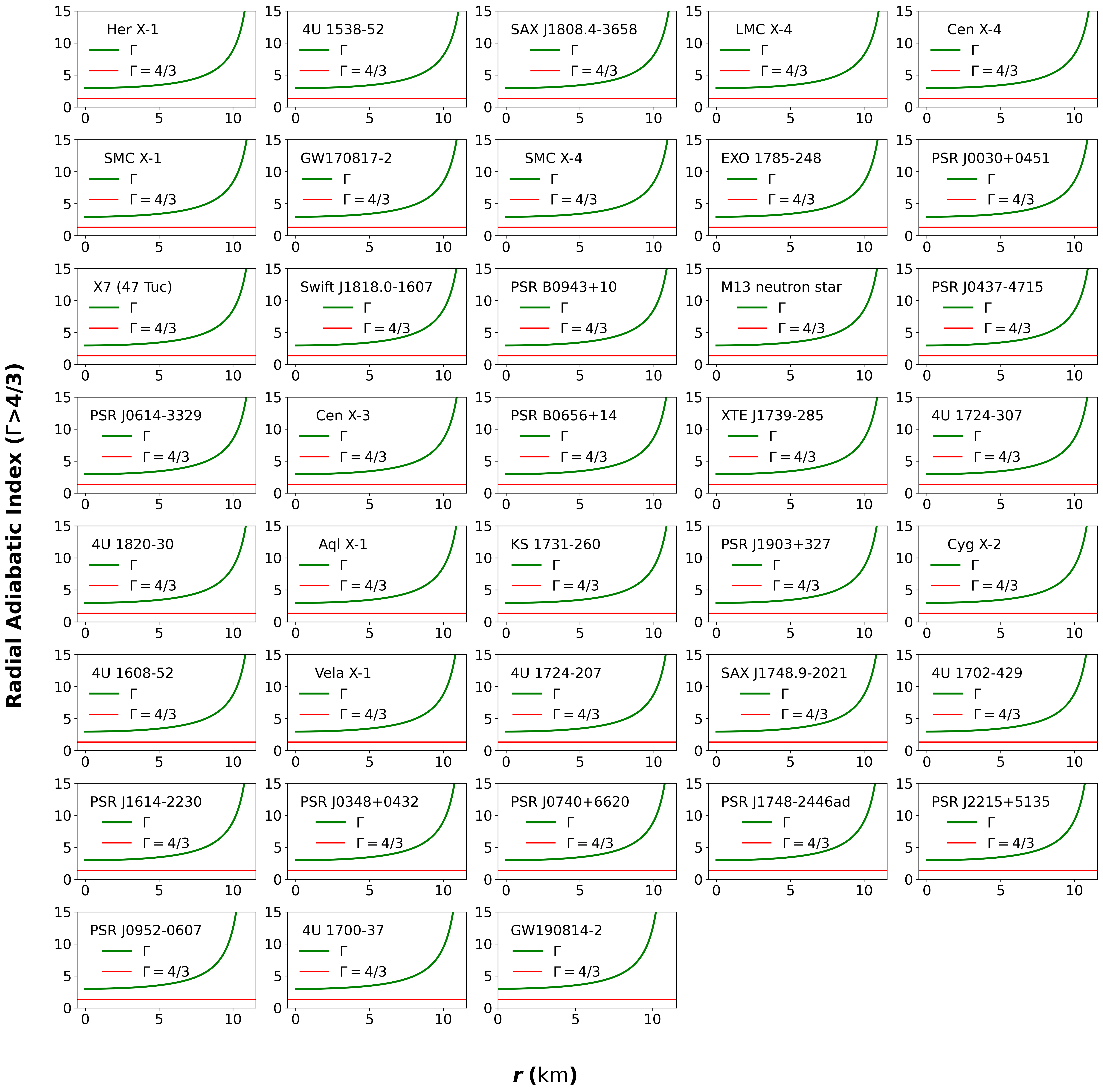}
\caption{illustrates the radial variation of the adiabatic index, $\Gamma(r)$ (green), together with the critical stability threshold, $\Gamma = 4/3$ (red), for the thirty-eight compact star configurations listed in Tables~\ref{tab:model_constants} and~\ref{tab:maxmassa}. The adiabatic index is evaluated throughout the stellar interior, extending from the centre ($r=0$) to the estimated stellar radius, $R_{\mathrm{Est}}$, which is determined by the physical boundary condition, $p_r(R_{\mathrm{Est}})=0$. Since the interior solution is physically valid only within the compact star boundary ($0 \leq r \leq R_{\mathrm{Est}}$), the profiles of $\Gamma(r)$ are presented exclusively over this interval. No extension of the solution into the exterior Schwarzschild space-time is considered, thereby ensuring that the stability analysis is confined to the physically meaningful interior region of each stellar configuration.}
\label{fig:adiabaticindex}
\end{center}
\end{figure}
\begin{itemize}
\item Each panel plots the radial adiabatic index $\Gamma(r)$ (green) against the reference
line $\Gamma = 4/3$ (red) across the stellar interior (Figure~\ref{fig:adiabaticindex}). The radial adiabatic index $\Gamma_r$ provides a standard indicator of dynamical-stability against the small radial perturbations, based on the Chandrasekhar stability criterion \cite{Chandrasekhar1964}. In the classical isotropic limit, the condition $\Gamma_r>4/3$ throughout the interior stellar configuration indicates stability against small radial perturbations, whereas a value of $\Gamma_r$ dropping below $4/3$ anywhere would indicate a dynamically unstable layer.

In every panel, the curve starts from a finite value near $r=0$ and rises sharply,
diverging as $r \to R_{Est}$. This divergence is expected from the surface boundary condition imposed on $p_r$ in Eq.~\eqref{eq10}. As $p_r \to 0$ near the stellar surface while the density, $\rho$, remains finite and nonzero, any quantity involving the ratio $(\rho+p_r)/p_r$ necessarily diverges as $r\to R_{Est}$. The rise observed across all thirty-eight panels is consistent with this structural feature of the model. However, the specific finite value attained by each curve near $r=0$ has not been independently verified against the corresponding fitted central density for each star, as these values are read directly from the plotted curves rather than calculated independently.

Across the plotted range, the green curve remains above the $\Gamma = 4/3$ reference
line in every panel, for all thirty-eight objects. Thus, $\Gamma>\frac{4}{3}$ throughout the stellar interior for all the neutron star configurations considered, thereby satisfying the well-established dynamical-stability criterion proposed by Chandrasekhar \cite{Chandrasekhar1964}. If this condition holds throughout the stellar interior, particularly near the center where the margin between $\Gamma$ and $4/3$ is typically smallest, it would indicate that the configurations satisfy the standard (isotropic-form) dynamical-stability criterion against radial perturbations throughout the interior. This is a necessary, though not sufficient, physical requirement for a viable stellar model, and should be confirmed numerically using the actual central density and pressure values for each star rather than relied upon from visual inspection of the plotted curves alone, particularly given how closely some curves may approach the reference line near $r=0$.

The radial adiabatic index is found to satisfy $\Gamma_r(r)>\frac{4}{3}$ throughout the stellar interior $0\leq r\leq R_{Est}$ for all thirty-eight configurations. This indicates that the configurations satisfy the standard adiabatic index criterion for stability. But for anisotropic relativistic stellar configurations,  the critical stability threshold may differ from $4/3$ due to the effects of relativistic gravity and pressure anisotropy ($p_t\neq p_r$). However, since the present analysis does not include a radial perturbation treatment of tangential pressure or equivalently an explicit anisotropy-dependent stability criterion, the condition $\Gamma_r>4/3$ should be regarded as a stability indicator rather than as a rigorous proof of dynamical-stability for the anisotropic stellar configurations.
\end{itemize}
\begin{figure}
\begin{center}
\includegraphics[width=0.9\textwidth,keepaspectratio]{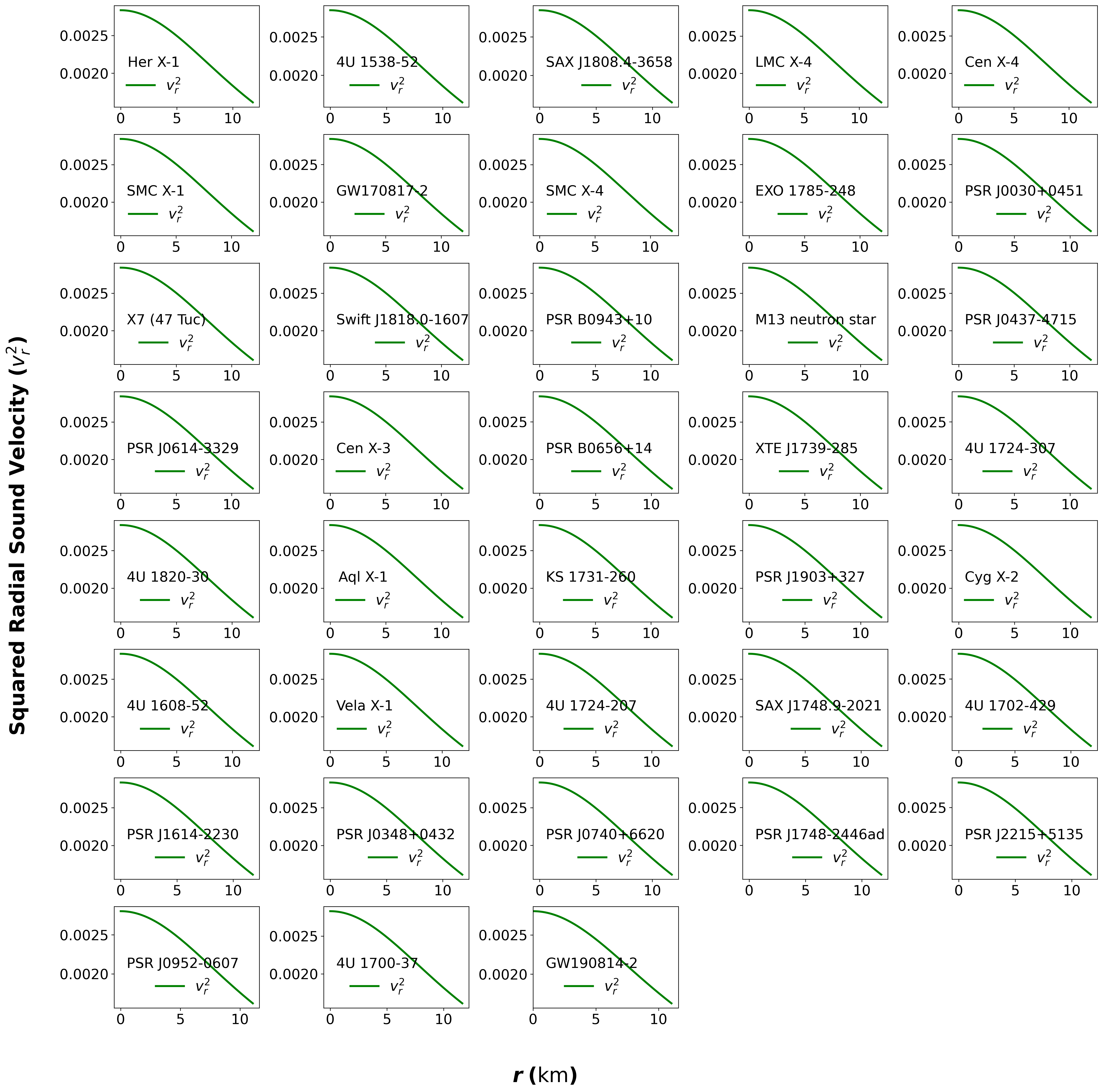}
\caption{illustrates the radial variation of the squared radial sound speed, $v_r^{2}(r)$, for the thirty-eight compact star configurations considered in the present study. The profiles are evaluated from the stellar centre ($r=0$) to the estimated stellar radius, $R_{\mathrm{Est}}$, determined by the physical boundary condition,  $p_r(R_{\mathrm{Est}})=0$. Since the interior solution is physically valid only within the compact star boundary ($0 \leq r \leq R_{\mathrm{Est}}$), the radial sound-speed profiles are presented exclusively over this interval. No extension into the exterior Schwarzschild space-time is considered, thereby ensuring that the causality analysis is restricted to the physically meaningful stellar interior. The obtained profiles satisfy the causality condition, $0 \leq v_r^{2} \leq 1$, throughout the interior of each compact star configuration, indicating that the propagation speed of radial perturbations does not exceed the speed of light.}
\label{fig:radialsound}
\end{center}
\end{figure}
\begin{itemize}
\item Each panel shows the squared radial sound speed, $v_r^2(r)=dp_r/d\rho$, in units where $c=1$, throughout the stellar interior for the same thirty-eight stellar configurations shown in Figure~\ref{fig:radialsound}. This provides the causality check for the model, since physical realizability of the adopted EoS requires the radial sound speed to satisfy $0\leq v_r^2\leq 1$ throughout the interior stellar configuration. This causality condition was imposed in the selection of the EoS parameters $X$ and $Y$.

In every panel, $v_r^2(r)$ is highest near the center and decreases
monotonically outward toward the surface of the stellar configurations, with no local minima, maxima, or non-monotonic features in any of the thirty-eight curves. This behavior is consistent with the corresponding density profiles shown in Figure~\ref{fig:rho}, where $\rho(r)$ decreases monotonically from the center toward the stellar surface. Since $dp_r/d\rho$ scales with a positive power of $\rho$ under the adopted polytropic-type EoS, the outward decrease in $\rho(r)$ naturally leads to a corresponding decrease in $v_r^2(r)$. Thus, the observed radial behavior of the squared radial sound speed is consistent with both the density profiles and the adopted EoS.

Across all thirty-eight panels, the plotted values of $v_r^2(r)$ fall within roughly $0.0020$--$0.0025$, comfortably inside the bound $0 \le v_r^2 \le 1$ over the visible axis range. Taken at face value, this is consistent with the matter remaining causal throughout the interior of every star. However, this is a weak confirmation because the plotted values sit so far below the causal limit that satisfying $v_r^2 \leq 1$ is essentially automatic from the axis scale and does not constitute a meaningful test of how close the model comes to the causal bound. A meaningful causality check would require the actual values relative to $1$, not just visual confirmation that the curves stay low on the plot.

These values are notably smaller than those typically reported for realistic neutron-star EoSs, for which $v_r^2/c^2$ can reach a substantial fraction of unity, often ranging from a few tenths to values of order unity at high central densities, rather than remaining confined to the $10^{-3}$ range throughout the stellar interior as found here. This could reflect the specific normalization or fitted value of $X$ used in this model, but that cannot be confirmed without the actual $X$ or central density $\rho_c$ for each star. Until this verification is performed, the magnitude of $v_r^2/c^2$ should be regarded as an open question rather than being dismissed as a scaling artifact or assumed to be physically expected. This is an important point to verify before interpreting the figure as evidence for causal and physically realistic stellar matter.

Using the fixed values of $X$ and the corresponding central density $\rho_c$ for each stellar configuration, the central radial sound speed was independently evaluated from $v_r^2(0)=\frac{3}{2}X\sqrt{\rho_c}$. The resulting values were checked against the causality condition $v_r^2/c^2\leq 1$ and found to remain within the allowed range for all configurations. Also the calculated values are in agreement with the corresponding values of $v_r^2$ at $r=0$ shown in Figure~\ref{fig:radialsound}, providing a direct numerical consistency check of the plotted profiles. This numerical verification confirms that the small values of $v_r^2/c^2$ at the stellar centre shown in Figure~\ref{fig:radialsound} are consistent with the adopted EoS and parameterization, rather than being an artifact of visual inspection.
\end{itemize}
\begin{figure}
\begin{center}
\includegraphics[width=0.9\textwidth,keepaspectratio]{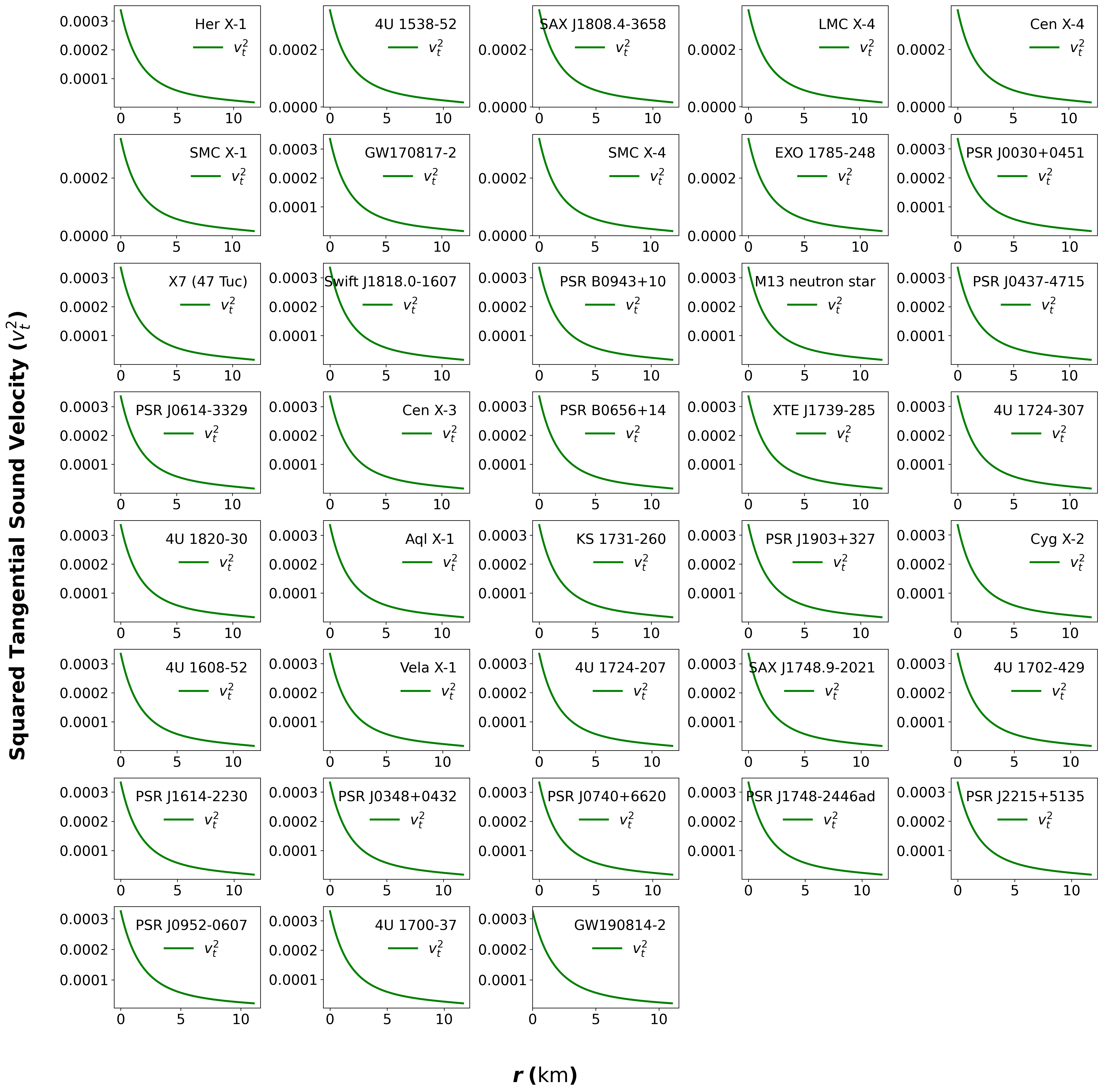}
\caption{illustrates the radial variation of the squared tangential sound speed, $v_t^{2}(r)$, for the thirty-eight compact star configurations considered in the present study within the framework of the proposed model. The profiles are evaluated from the stellar centre ($r=0$) to the estimated stellar radius, $R_{\mathrm{Est}}$, determined by the physical boundary condition, $p_r(R_{\mathrm{Est}})=0$. In conjunction with the corresponding radial sound-speed profiles shown in Fig.~\ref{fig:radialsound}, it is evident that the difference between the squared tangential and radial sound speeds satisfies $-1 \leq v_t^2-v_r^2 \leq 0$, throughout the stellar interior. Consequently, the configurations satisfy the Herrera cracking criterion \cite{Abreu2007}, indicating the absence of potentially unstable cracking regions and suggesting that the stellar models remain stable against local anisotropic perturbations. Since the interior solution is physically valid only within the compact star boundary ($0 \leq r \leq R_{\mathrm{Est}}$), the profiles of both the squared radial and tangential sound speeds, $v_r^{2}(r)$ and $v_t^{2}(r)$, are presented exclusively over this interval. No extension into the exterior Schwarzschild space-time is considered, thereby ensuring that the stability analysis is confined to the physically meaningful interior region of each compact star configuration.}
\label{fig:vt2}
\end{center}
\end{figure}
\begin{itemize}
\item Figure~\ref{fig:vt2} illustrates the radial variation of the squared tangential sound speed,
\begin{equation}
v_t^2=\frac{dp_t}{d\rho},
\end{equation}
for all the observed compact stars considered in the present investigation. It is evident that $v_t^2$ attains its maximum value at the stellar centre and decreases monotonically towards the stellar surface for every compact star candidate. This behaviour follows the corresponding decrease in both the tangential pressure and the energy density, indicating that the propagation of transverse pressure perturbations becomes progressively weaker away from the dense stellar core. Moreover, the smooth and continuous nature of the profiles demonstrates the absence of any discontinuity or singular behaviour throughout the stellar interior, thereby confirming the regularity of the proposed relativistic solution.

An important feature of Fig.~\ref{fig:vt2} is that the squared tangential sound speed remains positive throughout the stellar interior while satisfying the relativistic causality condition
\begin{equation}
0\leq v_t^2\leq 1.
\end{equation}
It is observed that the computed values are significantly smaller than unity over the entire radial domain, implying that the tangential propagation speed of pressure perturbations never exceeds the speed of light. Consequently, the proposed anisotropic compact star model remains fully compatible with the principle of causality and represents a physically admissible stellar configuration.

Furthermore, the gradual decline of $v_t^2$ from the centre towards the surface indicates that the effective tangential stiffness of the stellar matter is greatest in the high-density central region and decreases continuously with decreasing density. Since the adopted generalized polytropic EoS incorporates nonlinear density corrections, the resulting sound-speed profile naturally reflects the evolution of the thermodynamic response of dense-matter under extreme gravitational conditions. The absence of any abrupt variation in the profile also suggests that no unphysical behaviour or instability is introduced by the adopted model.

It is also noteworthy that all the observed compact stars exhibit an almost identical qualitative behaviour despite possessing different masses and radii. This consistency demonstrates that the proposed observation-constrained framework successfully reproduces physically acceptable tangential sound-speed distributions over a broad range of realistic compact star configurations. Therefore, the combined use of the Buchdahl space-time and the generalized polytropic EoS yields anisotropic stellar models that satisfy the fundamental requirements of causality and provide stable and physically realistic descriptions of the observed compact stars considered in the present study.
\end{itemize}
\begin{figure}
    \centering
    \includegraphics[width=0.9\textwidth,keepaspectratio]{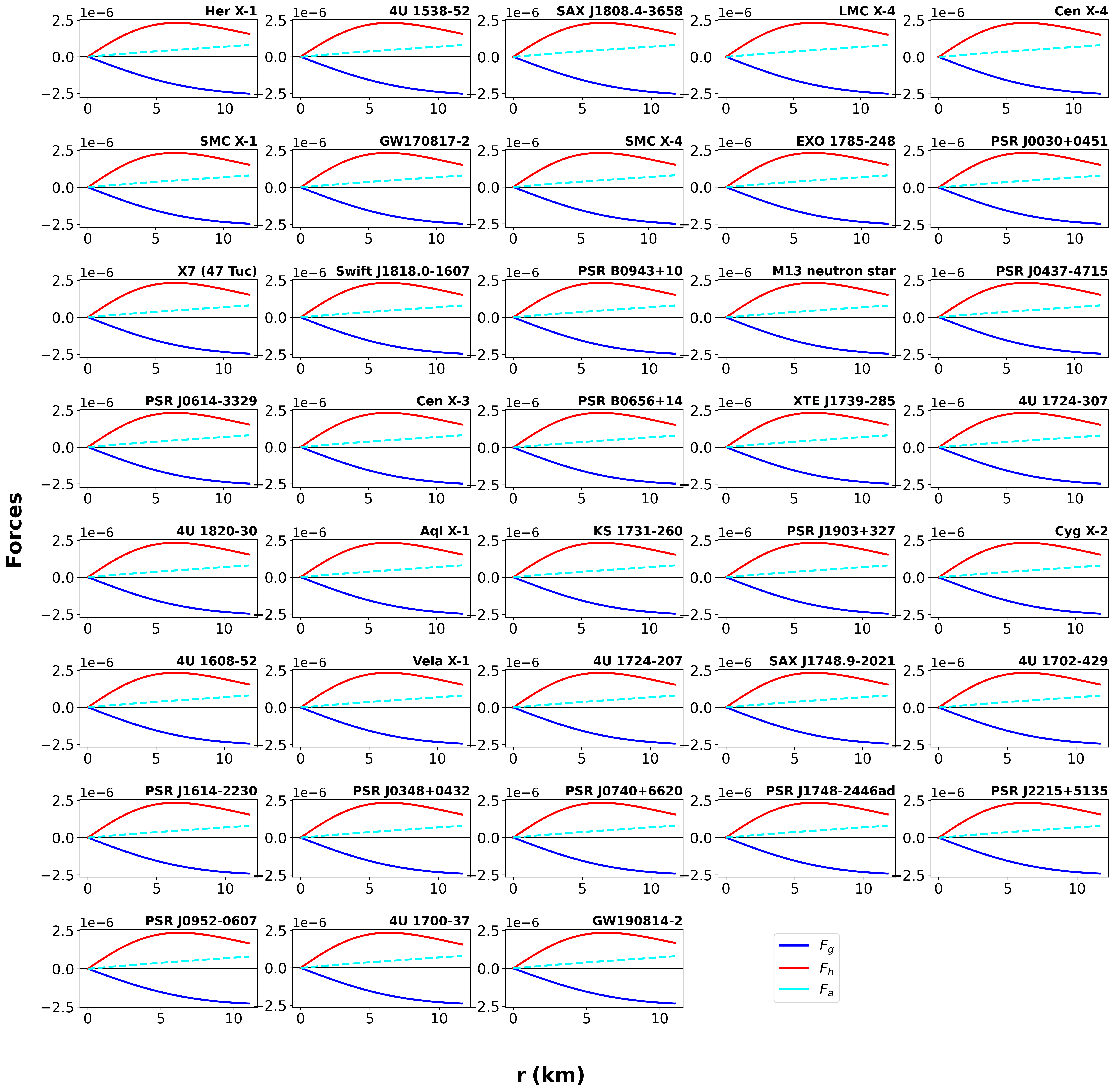}
    \caption{illustrates the radial variations of the individual force components appearing in the generalized TOV equilibrium equation for the thirty-eight compact star configurations considered in the present study. The gravitational force, $F_g$, hydrostatic force, $F_h$, and anisotropic force, $F_a$, are evaluated as functions of the radial coordinate, $r$, from the stellar centre ($r=0$) to the estimated stellar radius, $R_{\mathrm{Est}}$, determined by the physical boundary condition, $p_r(R_{\mathrm{Est}})=0$. The force profiles demonstrate that the inward-directed gravitational force is precisely counterbalanced by the combined outward contributions of the hydrostatic and anisotropic forces, thereby satisfying the generalized TOV equilibrium condition throughout the stellar interior. Moreover, the results indicate that the anisotropic force provides an outward contribution, which plays a significant role in supporting the compact configurations against gravitational collapse and maintaining their hydrostatic equilibrium. Since the interior solution is physically valid only within the compact star boundary ($0 \leq r \leq R_{\mathrm{Est}}$), the profiles of $F_g$, $F_h$, and $F_a$ are presented exclusively over this interval. No extension into the exterior Schwarzschild space-time is considered, ensuring that the equilibrium analysis is confined to the physically meaningful interior region of each stellar configuration.}
    \label{fig:TOV}
\end{figure}
\begin{itemize}
\item Figure~\ref{fig:TOV} illustrates the radial variation of the three forces governing the equilibrium of anisotropic compact stars, namely the gravitational force $F_g$, the hydrostatic force $F_h$, and the anisotropic force $F_a$, for all the observed compact star candidates considered in the present study. These forces satisfy the generalized TOV equations,
\begin{equation}
F_g+F_h+F_a=0,
\end{equation}
which represents the condition for hydrostatic equilibrium in a static spherically symmetric anisotropic stellar configuration. The individual force components appearing in the generalized TOV equilibrium equation can be expressed explicitly as \cite{Bowers1974,Herrera1997}
\begin{align}
F_g &= -\frac{1}{2}\left(\rho+p_r\right)\frac{d\alpha(r)}{dr},\qquad
F_h=-\frac{dp_r}{dr},\qquad
F_a=\frac{2}{r}\left(p_t-p_r\right).
\end{align}
 It is evident that all three forces vary smoothly throughout the stellar interior, indicating the absence of any discontinuity or singular behaviour within the proposed model.

The gravitational force $F_g$ remains negative over the entire radial domain, signifying its inward-directed nature arising from the self-gravity of the compact star. Its magnitude gradually increases from the stellar centre and then varies smoothly towards the surface, reflecting the cumulative gravitational attraction generated by the enclosed matter distribution. In contrast, the hydrostatic force $F_h$ is positive throughout the stellar interior and acts outward by opposing gravitational collapse through the pressure gradient. The close correspondence between the magnitudes of $F_g$ and $F_h$ demonstrates that the inward gravitational attraction is effectively balanced by the outward hydrostatic pressure.

The anisotropic force $F_a$ is also directed outward and remains positive throughout the stellar interior. Although its magnitude is comparatively smaller than those of the gravitational and hydrostatic forces, it provides an additional repulsive contribution that assists the pressure gradient in maintaining the equilibrium of the stellar configuration. Since the anisotropy vanishes at the stellar centre and gradually increases towards the surface, the corresponding anisotropic force exhibits a smooth radial increase, attaining its maximum value near the outer layers of the star. This behaviour is consistent with the physical expectation that anisotropic effects become increasingly important away from the central region.

It is noteworthy that all the observed compact stars exhibit nearly identical qualitative force distributions despite their different masses and radii. The inward gravitational force is precisely counterbalanced by the combined action of the hydrostatic and anisotropic forces throughout the stellar interior, thereby satisfying the generalized TOV equilibrium condition at every radial point. The excellent balance among these three forces confirms that the proposed observation-constrained model describes stable and physically realistic anisotropic compact star configurations within the framework of GR.
\end{itemize}
\begin{figure}
\begin{center}
\includegraphics[height=2.7in, width=3.4in]{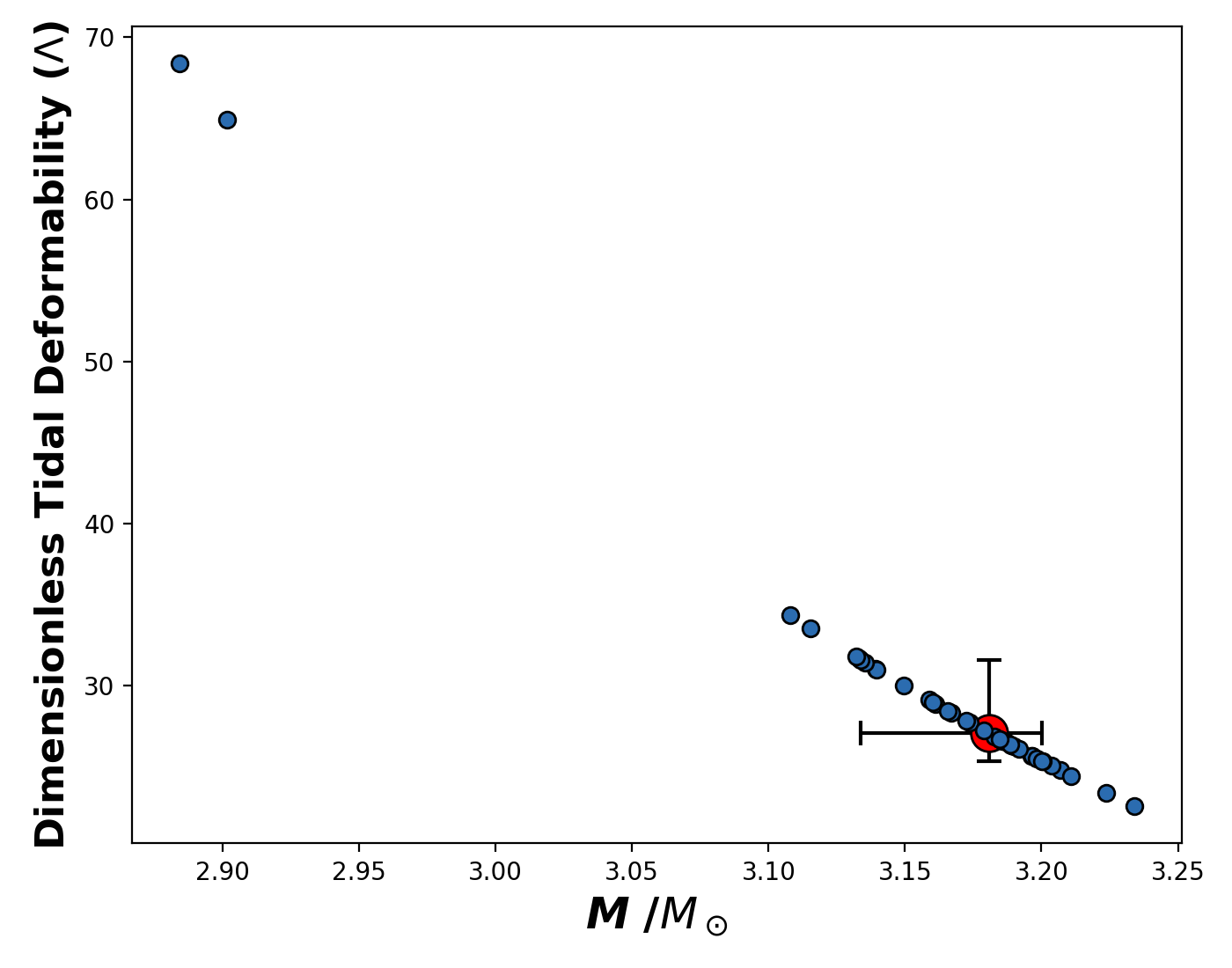}~~~~\includegraphics[height=2.67in, width=3.4in]{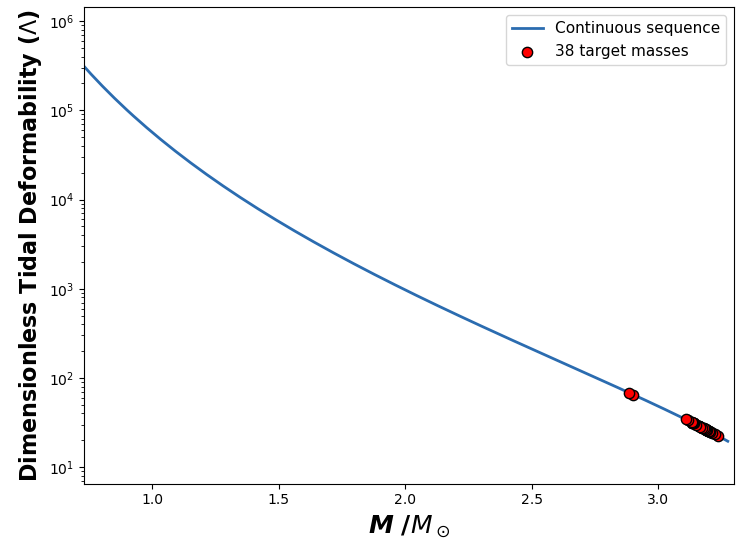}
\end{center}
\caption{The dimensionless tidal deformability, $\Lambda$, is presented as a function of the estimated stellar mass, $M_{\mathrm{Est}}$, for the thirty-eight considered stellar configurations. For each configuration, $\Lambda$ is evaluated at its corresponding estimated mass and radius., $M_{\mathrm{Est}}$ and $R_{\mathrm{Est}}$, respectively. The resulting set of $\Lambda$--$M_{\mathrm{Est}}$ data points is used to illustrate the variation of tidal deformability across the considered compact star models and to obtain an extrapolated trend. The median values of the estimated mass and tidal deformability are $M_{\mathrm{Est}}=3.1809^{+0.0194}_{-0.0471}\,\mathrm{M_{\odot}}$ and $\Lambda=27.04^{+4.54}_{-1.72}$, respectively (left panel). The calculated data points exhibit a clear inverse correlation between $M_{\mathrm{Est}}$ and $\Lambda$, with the tidal deformability progressively decreasing as the stellar mass increases. This behaviour indicates that more massive neutron stars are generally less susceptible to tidal deformation under an external tidal field. The extrapolated curve further extends this trend toward lower and higher stellar masses, showing an increase in $\Lambda$ toward lower masses and a decrease with increasing mass (Right panel). It should be emphasized, however, that the extrapolated portion represents an extension of the correlation established by the explicitly calculated stellar configurations and should not be interpreted as a direct prediction for stellar masses outside the range covered by the constructed models.
}
\label{fig:TD}
\end{figure}
\begin{itemize}
\item When a neutron star resides in a binary system with another compact object, the external tidal gravitational field generated by its companion perturbs the star from its equilibrium configuration, thereby inducing deviations from perfect spherical symmetry. Under the assumption that these tidal distortions are sufficiently small, the perturbations can be treated within the framework of linear perturbation theory. In this approach, the perturbed space-time is described by introducing a small metric perturbation, $\eta_{\mu\nu}$, to the static background interior metric tensor, $g^{-}_{\mu\nu}$. Accordingly, the interior space-time metric of the tidally deformed compact star may be written as \cite{Hinderer2008}
$$(g^-_{\mu\nu})_{modified}=g^-_{\mu\nu}+\eta_{\mu\nu}~~~,$$
represents the perturbed space-time. In this regard the small perturbation $\eta_{\mu\nu}$ is of the form  \cite{Binnington2009}
\[
\eta_{\mu\nu} =
\begin{pmatrix}
e^{p(r)} \mathcal{H}(r) & 0 & 0 & 0 \\
0 & e^{q(r)} \mathcal{H}(r) & 0 & 0 \\
0 & 0 & r^2 \mathcal{K}(r) & 0 \\
0 & 0 & 0 & r^2 \sin^2\theta \, \mathcal{K}(r)
\end{pmatrix}
P_2(\cos\theta)~~~~.
\]
Here, $\mathcal{H}(r)$ and $\mathcal{K}(r)$ denote the radial metric perturbation functions associated with the even-parity perturbations of the static, spherically symmetric background space-time, while $P_{2}(\cos\theta)$ represents the Legendre polynomial corresponding to the quadrupolar mode ($l=2$). Restricting the analysis to static, axisymmetric quadrupolar perturbations, the interior space-time metric given by Eq.~\eqref{eq3} assumes the perturbed form \cite{Hinderer2008,Binnington2009}
\[
ds_{-}^2 =
- e^{p(r)}\left[1 + \mathcal{H}(r) P_2(\cos\theta)\right] dt^2
+ e^{q(r)}\left[1 + \mathcal{H}(r) P_2(\cos\theta)\right] dr^2
+ r^2 \left[1 + \mathcal{K}(r) P_2(\cos\theta)\right]
\left(d\theta^2 + \sin^2\theta d\phi^2\right)~~~.
\]
Furthermore, the radial perturbation function $\mathcal{K}(r)$ is not independent and can be expressed in terms of the perturbation function $\mathcal{H}(r)$ and its first radial derivative, i.e., $\mathcal{K}(r)=f\!\left(\mathcal{H}(r),\mathcal{H}'(r)\right)$. Substituting this relation into the perturbed Einstein field equations and eliminating $\mathcal{K}(r)$, one obtains a second-order ordinary differential equation governing the radial perturbation function $\mathcal{H}(r)$, which can be written in the generic form \cite{Hinderer2008}
\begin{equation}\label{chal1}
r\frac{dy(r)}{dr}+y^2+y(r)\mathcal{F}(r)+r^2\mathcal{Q}(r)=0~~~,
\end{equation}
where $y(r)=r\mathcal{H}'(r)/\mathcal{H}(r)$. Moreover, the functions $\mathcal{F}(r)$ and $\mathcal{Q}(r)$ depend explicitly on the radial distributions of the energy density $\rho(r)$, the radial pressure $p_r(r)$ and the enclosed gravitational mass, $M(r)$, of the compact stellar configuration. These functions can be expressed as \cite{Hinderer2008,Binnington2009} 
\[
\mathcal{F}(r)=\frac{r-4\pi r^3\left\{\rho(r)-p_r(r)\right\}}{r-2M(r)}~~~\text{and}\]
\[
\mathcal{Q}(r)=\frac{4\pi r\left\{5\rho(r)+9p_r(r)+
\frac{\rho(r)+p_r(r)}{v_r^2}\right\}}{\left\{r-2M(r)\right\}}
-\frac{6}{r^2}
-\frac{4\left\{M(r)+4\pi r^3 p_r(r)\right\}^2}{r^2\left\{r-2M(r)\right\}^2}~~~.
\]
Here, the squared radial sound speed is defined as $v_r^{2}=dp_r/d\rho$. By integrating the differential equation \eqref{chal1} from the stellar centre ($r=0$) to the stellar surface ($r=R_{Est}$), subject to the regularity condition at the centre, the surface value of the perturbation function, denoted by $y_F \equiv y(R_{Est})$, is obtained. The dimensionless tidal Love number, $\kappa$, which characterizes the quadrupolar tidal response of the compact star, can then be expressed as a function of the stellar compactness $u(r)=M(r)/{r}$, and the quantity $y_F$. The corresponding expression for the tidal Love number is given by
$$\kappa =\frac{8u^5}{5}\Bigg(1-2u\Bigg)^2\Bigg\{2u(y_F-1)-y_F+2\Bigg\}\Bigg[2u\Bigg\{6-3y_F+3u(5y_F-8)\Bigg\}
+4u^3\Bigg\{13-11y_F+u(3y_F-2)$$
\begin{equation}\label{arpann}
~~~~~~~~~~~~~~~~~~~~~~~~~~~~~~~~~~~~~~~~~~~~~~~~~~~~~~+2u^2(1+y_F)\Bigg\}+3(1-2u)^2\Bigg\{2-y_F+2u(y_F-1)\Bigg\}\ln(1-2u)\Bigg]^{-1}.
\end{equation}
Now with help of $\kappa$, tidal deformability ($\Lambda$) of the corresponding stellar structure can be depicted as
$$\Lambda=\frac{2}{3}\kappa u^{-5}~~~.$$

The dimensionless tidal deformability, $\Lambda$, provides a measure of the response of a neutron star to an external tidal gravitational field. The variation shown in Figure~\ref{fig:TD} exhibits a clear inverse correlation between the estimated stellar mass, $M_{\mathrm{Est}}$ and $\Lambda$. As $M_{\mathrm{Est}}$ increases from approximately $2.88\,\mathrm{M_{\odot}}$ to $3.23\,\mathrm{M_{\odot}}$, $\Lambda$ decreases from about $68.38$ to $22.55$. The median values are obtained as $M_{\mathrm{Est}}=3.1809^{+0.0194}_{-0.0471}\,\mathrm{M_{\odot}}$ and $\Lambda=27.04^{+4.54}_{-1.72}$, indicating that the considered configurations generally correspond to highly compact stellar objects with relatively small tidal responses. Physically, the decrease in $\Lambda$ with increasing mass is associated with the increasing compactness of the stellar configurations. A more massive and compact neutron star is more strongly bound by its gravitational field and therefore undergoes less deformation under an external tidal field. Consequently, the lower values of $\Lambda$ indicate greater resistance to tidal distortion.

The binary neutron star merger event GW170817 provides an upper constraint on the dimensionless tidal deformability of a $1.4\,\mathrm{M_{\odot}}$ neutron star, $\Lambda_{1.4}\lesssim800$ \cite{Abbott2017,Abbott2018}. The stellar configurations considered in the present research work span the estimated mass range $2.88\lesssim M_{\mathrm{Est}}/M_{\odot}\lesssim3.23$, with the corresponding dimensionless tidal deformabilities lying within the range $22.55\lesssim\Lambda\lesssim68.38$. Although the present models do not directly represent a $1.4\,\mathrm{M_{\odot}}$ neutron star, all the predicted values of $\Lambda$ remain substantially below the upper bound inferred from GW170817. This behavior indicates a comparatively weak tidal response and strong resistance to deformation for the massive compact configurations considered here. The observed monotonic decrease of $\Lambda$ with increasing mass therefore provides a physically consistent indication of increasing stellar compactness and gravitational binding within the proposed model.

Moreover these calculated $\Lambda$--$M_{\rm Est}$ data points exhibit a clear inverse correlation between tidal deformability and stellar mass. An extrapolation of this correlation toward lower masses indicates a
rapid increase in $\Lambda$, reaching values of order $10^5$ near $M\simeq0.83\,M_{\odot}$. This value should be regarded as an extrapolated estimate rather than a direct prediction, since the
stellar configurations explicitly constructed in the present analysis do not extend to this mass. The observational constraint $\Lambda_{1.4}\lesssim800$ inferred from GW170817 is relevant to the
configuration at approximately $1.4\,M_{\odot}$ and therefore should not be directly applied to the extrapolated low-mass value.
\end{itemize}
\begin{figure}
\includegraphics[width=0.9\textwidth,keepaspectratio]{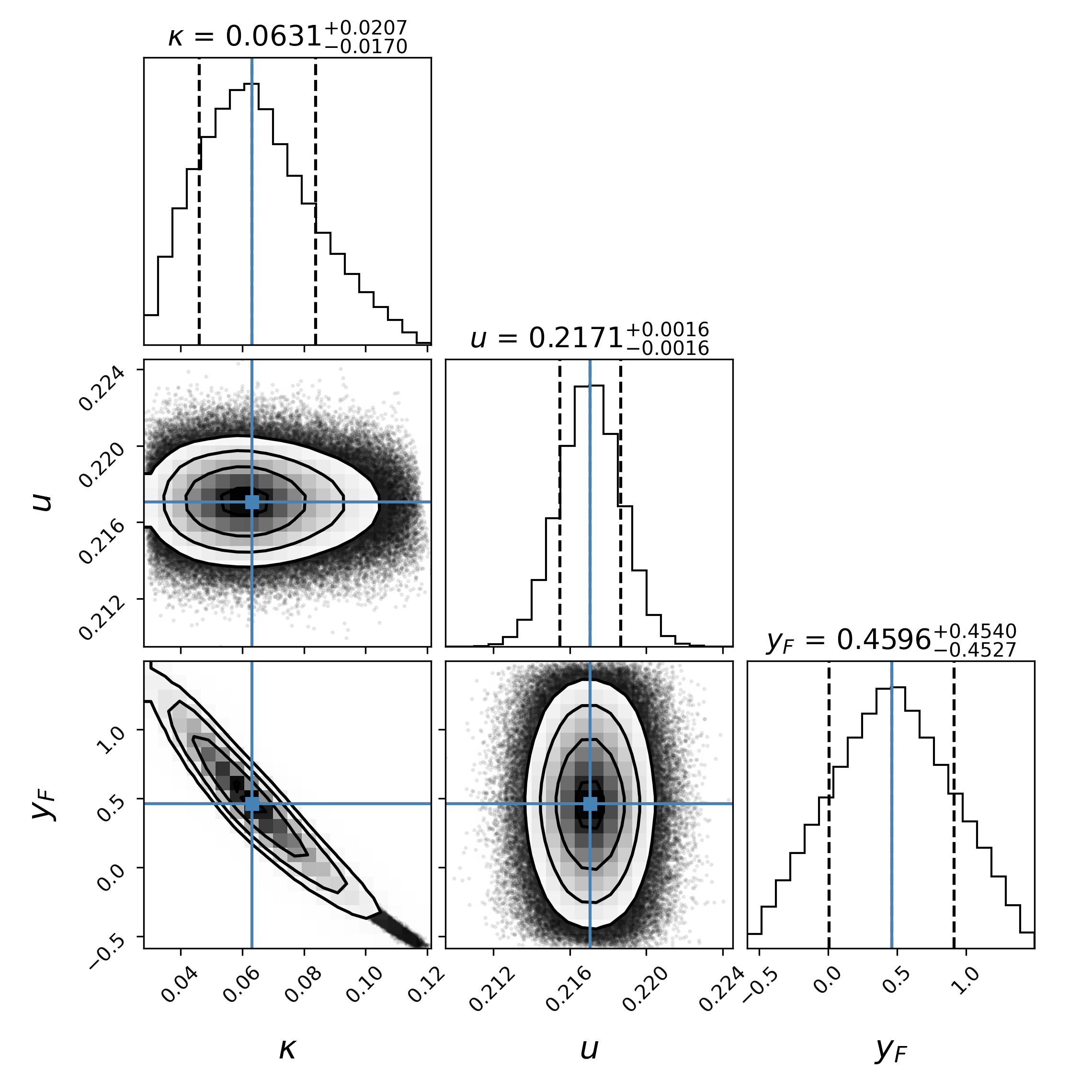}
\caption{compactness parameter is defined as $u_{\mathrm{obs}}=M_{\mathrm{obs}}\,M_{\odot}/R_{\mathrm{obs}}$, and its associated uncertainty is given by $\sigma_u=\sqrt{\left(\sigma_M/M_{\mathrm{obs}}\right)^2+\left(\sigma_R/R_{\mathrm{obs}}\right)^2}$. The fitting function is assumed to follow a linear relation of the form $y_F=a-bu$, where $a$ and $b$ are free parameters. The parameter $\kappa$ is obtained from Eq.~\eqref{arpann}. The priors are chosen as $0.05<u<0.35$, $0.5<a<1.5$ and $0.0<b<5.0$. Parameter estimation is performed using a chi-square minimization approach, with sampling carried out via the Markov Chain Monte Carlo (MCMC) method. The best-fit values correspond to the median ($50^{\mathrm{th}}$ percentile) of the posterior distributions, while uncertainties are determined from the $1\sigma$ (68\%) credible intervals. Contour levels corresponding to $1\sigma$, $2\sigma$, and $3\sigma$ confidence regions are also shown. The acceptance fraction was approximately $0.537828^{+0.005412}_{-0.007888}$~. The MCMC chains satisfied the convergence criterion ($\hat{R}<1.01$).}
\label{fig:MCMC1}
\end{figure}
\begin{itemize}
\item Figure~\ref{fig:MCMC1} presents the posterior distributions and parameter correlations obtained from the MCMC analysis. The compactness parameter is defined as
\begin{equation}
u=\frac{M_{\mathrm{obs}}M_{\odot}}{R_{\mathrm{obs}}},
\end{equation}
and its median value is obtained as
\begin{equation}
u=0.2171^{+0.0016}_{-0.0016}.
\end{equation}
This value indicates that the considered stellar configurations are highly compact and possess strong gravitational fields at their surfaces. Physically, a larger compactness corresponds to stronger gravitational binding and a more significant departure from the Newtonian regime.

The parameter $\kappa$ is obtained as
\begin{equation}
\kappa=0.0631^{+0.0207}_{-0.0170},
\end{equation}
which characterizes the model-dependent contribution associated with the compact stellar configuration. The relatively localized posterior distribution of $\kappa$ indicates that this parameter is reasonably well constrained by the adopted observational data and fitting procedure.

The fitted parameter $y_F$ is obtained as
\begin{equation}
y_F=0.4596^{+0.4540}_{-0.4527}.
\end{equation}
The broad posterior distribution of $y_F$ indicates a comparatively large uncertainty in this parameter. A strong anticorrelation is observed between $\kappa$ and $y_F$, as shown by the elongated contours in the $\kappa$--$y_F$ plane. This behavior indicates that an increase in one parameter is accompanied by a decrease in the other, while a comparable fit to the observational data is maintained. Therefore, $\kappa$ and $y_F$ are not independently constrained with equal strength.

The posterior distribution of $u$ is comparatively narrow, indicating that the compactness is more tightly constrained than $y_F$. The approximately elliptical posterior contours indicate a well-defined preferred region in the parameter space. The $1\sigma$, $2\sigma$, and $3\sigma$ contours represent the corresponding confidence regions and illustrate the uncertainties associated with the inferred parameters. The convergence criterion,
\begin{equation}
\hat{R}<1.01,
\end{equation}
indicates satisfactory convergence of the MCMC chains. Hence, the obtained posterior distributions are considered statistically reliable within the adopted model and priors.

Overall, the MCMC analysis indicates that the observational data favor compact stellar configurations with $u\simeq0.217$. The parameter correlations further indicate that the model parameters can partially compensate for one another while reproducing the observed stellar properties. The inferred compactness therefore provides evidence for a strongly gravitating stellar configuration. In contrast, the broader uncertainty in $y_F$ indicates that additional observational constraints are required for a more precise determination of this parameter.
\end{itemize}
\section{Brief Discussion and Conclusion}

In this work, an anisotropic compact star model is developed within the framework of GR by combining the Buchdahl interior space-time with the adopted generalized polytropic EoS. The model parameters are constrained through the Darmois--Israel matching conditions by smoothly joining the interior geometry to the exterior Schwarzschild space-time. The resulting framework is applied to thirty-eight observed compact star candidates spanning a broad range of masses and radii. The model successfully reproduces physically acceptable compact stellar configurations and provides a unified description of the considered observational sample. The estimated compactness increases systematically with stellar mass and remains below the classical Buchdahl limit used in the present analysis as a consistency check, although the model itself is anisotropic.

The interior matter variables exhibit regular and physically meaningful behaviour for all the considered stellar configurations. The energy density remains finite and positive at the centre and decreases monotonically towards the surface, whereas the enclosed gravitational mass increases continuously from the centre and reaches its maximum value at the stellar boundary. The radial pressure decreases outward and vanishes at the estimated stellar radius, thereby defining the physical boundary of each configuration. The tangential pressure also remains regular throughout the interior. The difference between the tangential and radial pressures generates anisotropy, which vanishes at the stellar centre and contributes to the internal force balance. The obtained configurations also satisfy the Zeldovich condition, supporting the physical acceptability of the adopted matter distribution.

The gravitational redshift remains finite and increases monotonically from the stellar centre to the surface, with its terminal value depending on the compactness of the corresponding configuration. More compact objects therefore exhibit stronger surface gravitational effects. The dynamical-stability analysis further shows that the radial adiabatic index remains above the critical value $\Gamma=4/3$ throughout the stellar interior for all thirty-eight configurations, indicating stability against radial adiabatic perturbations within the adopted criterion. Both the radial and tangential squared sound speeds remain positive and satisfy the causality conditions $0\leq v_r^2\leq1$ and $0\leq v_t^2\leq1$, respectively. Moreover, the difference between the two sound speeds satisfies the Herrera cracking condition, suggesting the absence of potentially unstable cracking regions and supporting the stability of the anisotropic configurations against local perturbations.

The generalized Tolman--Oppenheimer--Volkoff equilibrium condition is also satisfied throughout the stellar interior. The inward-directed gravitational force is balanced by the combined outward hydrostatic and anisotropic forces. Although the anisotropic contribution is smaller than the gravitational and hydrostatic components, its outward nature provides additional support against gravitational collapse and becomes particularly relevant in maintaining the equilibrium of the compact configurations. The smooth behaviour of all three force components confirms the regularity and hydrostatic stability of the proposed stellar models.

The tidal analysis reveals a clear inverse correlation between the estimated stellar mass and the dimensionless tidal deformability. For the constructed configurations, the estimated mass spans approximately $2.88\lesssim M_{\rm Est}/M_\odot\lesssim3.23$, while the corresponding tidal deformability lies within $22.55\lesssim\Lambda\lesssim68.38$. Thus, more massive and compact configurations exhibit smaller tidal deformabilities and are more resistant to external tidal distortion. The median values are obtained as $M_{\rm Est}=3.1809^{+0.0194}_{-0.0471}M_\odot$ and $\Lambda=27.04^{+4.54}_{-1.72}$. The predicted tidal deformabilities remain substantially below the upper constraint associated with GW170817, although the present models do not directly represent a $1.4\,M_\odot$ neutron star. The extrapolated low-mass trend should, however, be interpreted only as an extension of the correlation obtained from the explicitly constructed configurations and not as a direct prediction outside the modelled mass range.

The statistical analysis based on the MCMC method further supports the observational consistency of the model. The compactness is obtained as $u=0.2171^{+0.0016}_{-0.0016}$, indicating strongly gravitating compact configurations, while the tidal Love number parameter is found to be $\kappa=0.0631^{+0.0207}_{-0.0170}$. The surface perturbation parameter is obtained as $y_F=0.4596^{+0.4540}_{-0.4527}$, with its broader posterior distribution indicating comparatively larger uncertainty. A strong anticorrelation between $\kappa$ and $y_F$ is observed, showing that these parameters can partially compensate for one another while maintaining consistency with the observational constraints. The narrow posterior distribution of compactness and the convergence criterion $\hat{R}<1.01$ demonstrate satisfactory convergence and statistical reliability within the adopted model and priors.

Overall, the present investigation demonstrates that the Buchdahl space-time coupled with the adopted generalized polytropic EoS provides a mathematically consistent and physically viable framework for describing anisotropic compact stars over a broad observational range. The simultaneous satisfaction of regularity, positive and decreasing density and pressure profiles, physically acceptable anisotropy, finite redshift, causality, radial stability, cracking stability, hydrostatic equilibrium, and physically consistent tidal behaviour supports the viability of the constructed stellar configurations. The results further indicate that anisotropy can provide an additional outward force capable of supporting highly compact and massive stellar objects. The model therefore offers a useful phenomenological framework for investigating the internal structure and tidal response of compact stars, while future studies may further constrain the equation-of-state parameters using multimessenger observations and extend the analysis to rotating configurations and more realistic dense-matter descriptions.
\section*{Acknowledgments}
RB gratefully acknowledges the Inter-University Centre for Astronomy and Astrophysics (IUCAA), Pune, India, for the Visiting Associateship. AB gratefully acknowledges the financial support received through the UGC-NET Senior Research Fellowship (SRF). SP gratefully acknowledges the financial support received through the Swami Vivekananda Merit-cum-Means Scholarship for Ph.D. Research Fellowship, Higher Education Department, Government of West Bengal.
\section*{Funding}
There is no funding to report for this article.
\section*{Data Availability Statement}
This article has no associated data as the work is completely theoretical in nature.
\section*{Code Availability Statement}
My manuscript has no associated code/software.
\section*{Conflict of Interest}
The authors declare that they have no known competing financial interests or personal relationships that could have appeared to influence the work reported in this paper. 

\bibliographystyle{ieeetr} 

\bibliography{bibliography}

@article{Miller2019,
  author = {Miller, M. C. et al.},
  title = {PSR J0030+0451 Mass and Radius from NICER Data},
  journal = {Astrophys. J. Lett.},
  volume = {887},
  pages = {L24},
  year = {2019}
}

@article{Riley2019,
  author = {Riley, T. E. et al.},
  title = {A NICER View of PSR J0030+0451},
  journal = {Astrophys. J. Lett.},
  volume = {887},
  pages = {L21},
  year = {2019}
}

@article{Miller2021,
  author = {Miller, M. C. et al.},
  title = {PSR J0740+6620 Mass and Radius Constraints},
  journal = {Astrophys. J. Lett.},
  volume = {918},
  pages = {L28},
  year = {2021}
}

@article{Riley2021,
  author = {Riley, T. E. et al.},
  title = {NICER Constraints on PSR J0740+6620},
  journal = {Astrophys. J. Lett.},
  volume = {918},
  pages = {L27},
  year = {2021}
}

@article{Abbott2017,
  author = {Abbott, B. P. et al.},
  title = {GW170817: Observation of Gravitational Waves from a Binary Neutron Star Inspiral},
  journal = {Phys. Rev. Lett.},
  volume = {119},
  pages = {161101},
  year = {2017}
}

@article{Abbott2018,
  author = {Abbott, B. P. et al.},
  title = {GW170817: Measurements of Neutron Star Radii and EOS},
  journal = {Phys. Rev. Lett.},
  volume = {121},
  pages = {161101},
  year = {2018}
}

@article{Tolman1939,
  author  = {R. C. Tolman},
  title   = {Static Solutions of Einstein's Field Equations for Spheres of Fluid},
  journal = {Physical Review},
  volume  = {55},
  pages   = {364--373},
  year    = {1939}
}

@article{Oppenheimer1939,
  author  = {J. R. Oppenheimer and G. M. Volkoff},
  title   = {On Massive Neutron Cores},
  journal = {Physical Review},
  volume  = {55},
  pages   = {374--381},
  year    = {1939}
}

@article{Schwarzschild1916,
  author  = {K. Schwarzschild},
  title   = {On the Gravitational Field of a Mass Point according to Einstein's Theory},
  journal = {Sitzungsberichte der K{\"o}niglich Preussischen Akademie der Wissenschaften},
  pages   = {189--196},
  year    = {1916}
}

@article{Darmois1927,
  author  = {G. Darmois},
  title   = {Les \'equations de la gravitation einsteinienne},
  journal = {M\'emorial des sciences math\'ematiques},
  volume  = {25},
  pages   = {1--25},
  year    = {1927}
}

@article{Israel1966,
  author  = {W. Israel},
  title   = {Singular hypersurfaces and thin shells in general relativity},
  journal = {Il Nuovo Cimento B},
  volume  = {44},
  pages   = {1--14},
  year    = {1966}
}

@article{Cromartie2020,
  author  = {H. T. Cromartie and et al.},
  title   = {Relativistic Shapiro delay measurements of an extremely massive millisecond pulsar},
  journal = {Nature Astronomy},
  volume  = {4},
  pages   = {72--76},
  year    = {2020}
}

@article{Lattimer2020,
  author  = {J. M. Lattimer and M. Prakash},
  title   = {Neutron Star Structure and the Equation of State},
  journal = {Astrophysical Journal},
  volume  = {550},
  pages   = {426},
  year    = {2000}
}

@book{Glendenning2000,
  author    = {N. K. Glendenning},
  title     = {Compact Stars: Nuclear Physics, Particle Physics and General Relativity},
  publisher = {Springer},
  year      = {2000}
}

@book{Haensel2007,
  author    = {P. Haensel and A. Y. Potekhin and D. G. Yakovlev},
  title     = {Neutron Stars 1: Equation of State and Structure},
  publisher = {Springer},
  year      = {2007}
}

@article{Lattimer2016,
  author  = {J. M. Lattimer and M. Prakash},
  title   = {The Equation of State of Hot, Dense Matter},
  journal = {Physics Reports},
  volume  = {621},
  pages   = {127--164},
  year    = {2016}
}

@article{Chavanis2015,
  author  = {P. H. Chavanis},
  title   = {Cosmology with a stiff matter era},
  journal = {European Physical Journal C},
  volume  = {75},
  pages   = {130},
  year    = {2015}
}

@article{Demorest2010,
  author  = {Demorest, P. B. and et al.},
  title   = {A two-solar-mass neutron star measured using Shapiro delay},
  journal = {Nature},
  volume  = {467},
  pages   = {1081--1083},
  year    = {2010}
}

@article{Maurya2017,
  author  = {Maurya, S. K.},
  title   = {A new class of relativistic charged anisotropic solution for compact stars},
  journal = {European Physical Journal C},
  volume  = {77},
  pages   = {45},
  year    = {2017}
}

@article{Sharma2020,
  author  = {Sharma, R. and Maharaj, S. D.},
  title   = {A class of relativistic stars with a linear equation of state},
  journal = {European Physical Journal C},
  volume  = {80},
  pages   = {39},
  year    = {2020}
}

@article{Sharma2007,
  author  = {Sharma, R. and Maharaj, S. D.},
  title   = {A class of relativistic stars with a linear equation of state},
  journal = {Monthly Notices of the Royal Astronomical Society},
  volume  = {375},
  number  = {4},
  pages   = {1265--1268},
  year    = {2007},
  doi     = {10.1111/j.1365-2966.2006.11355.x}
}

@article{Deb2021,
  author  = {Deb, D. and et al.},
  title   = {Exploring physical constraints on compact stars beyond the neutron star maximum mass},
  journal = {Monthly Notices of the Royal Astronomical Society},
  volume  = {507},
  pages   = {1583--1595},
  year    = {2021}
}

@article{Mauviard2025,
  author  = {Mauviard, L. and et al.},
  title   = {Constraints on neutron star structure from multimessenger observations},
  journal = {Astrophysical Journal},
  volume  = {995},
  number  = {1},
  pages   = {60},
  year    = {2025}
}

@article{Choudhury2024,
  author  = {Choudhury, D. and et al.},
  title   = {Evidence for dense matter effects in neutron star mergers from GW observations},
  journal = {Astrophysical Journal Letters},
  volume  = {971},
  number  = {1},
  pages   = {L20},
  year    = {2024}
}

@article{Bowers1974,
  author  = {Bowers, R. L. and Liang, E. P. T.},
  title   = {Anisotropic spheres in general relativity},
  journal = {Astrophysical Journal},
  volume  = {188},
  pages   = {657--665},
  year    = {1974}
}

@article{Herrera1997,
  author  = {Herrera, L. and Santos, N. O.},
  title   = {Local anisotropy in self-gravitating systems},
  journal = {Physics Reports},
  volume  = {286},
  pages   = {53--130},
  year    = {1997}
}

@article{Hinderer2008,
  author  = {Hinderer, T.},
  title   = {Tidal Love numbers of neutron stars},
  journal = {Astrophysical Journal},
  volume  = {677},
  pages   = {1216--1220},
  year    = {2008}
}

@article{Binnington2009,
  author  = {Binnington, T. and Poisson, E.},
  title   = {Relativistic theory of tidal Love numbers},
  journal = {Physical Review D},
  volume  = {80},
  pages   = {084018},
  year    = {2009}
}

@article{Abreu2007,
  author  = {Abreu, H. and Hern{\'a}ndez, H. and N{\'u}{\~n}ez, L. A.},
  title   = {Stability of anisotropic spheres in general relativity},
  journal = {Classical and Quantum Gravity},
  volume  = {24},
  number  = {18},
  pages   = {4631--4646},
  year    = {2007},
  doi     = {10.1088/0264-9381/24/18/012}
}

@article{Ivanov2002,
  author = {Ivanov, B. V.},
  title = {Maximum bounds on the surface redshift of anisotropic stars},
  journal = {Physical Review D},
  volume = {65},
  pages = {104011},
  year = {2002}
}

@article{Buchdahl1959,
  author = {Buchdahl, H. A.},
  title = {General Relativistic Fluid Spheres},
  journal = {Physical Review},
  volume = {116},
  pages = {1027--1034},
  year = {1959}
}

@article{Chandrasekhar1964,
  author = {Chandrasekhar, S.},
  title = {Dynamical Instability of Gaseous Masses Approaching the Schwarzschild Limit},
  journal = {Astrophysical Journal},
  volume = {140},
  pages = {417--433},
  year = {1964}
}

@article{Rawls2011,
  author  = {Rawls, M. L. and et al.},
  title   = {Refined Neutron-Star Mass Determinations for Six Eclipsing X-Ray Pulsar Binaries},
  journal = {Astrophysical Journal},
  volume  = {730},
  pages   = {25},
  year    = {2011}
}

@article{Ozel2009,
  author  = {\"Ozel, F. and et al.},
  title   = {The Mass and Radius of the Neutron Star in EXO 1745-248},
  journal = {Astrophysical Journal},
  volume  = {693},
  pages   = {1775},
  year    = {2009}
}

@article{Antoniadis2013,
  author  = {Antoniadis, J. and et al.},
  title   = {A Massive Pulsar in a Compact Relativistic Binary},
  journal = {Science},
  volume  = {340},
  pages   = {448},
  year    = {2013}
}

@article{Romani2022,
  author  = {Romani, R. W. and et al.},
  title   = {PSR J0952-0607: The Fastest and Most Massive Known Galactic Neutron Star},
  journal = {Astrophysical Journal Letters},
  volume  = {934},
  pages   = {L17},
  year    = {2022}
}

@article{delaCruzDombriz2009,
  author  = {de la Cruz-Dombriz, A. and Dobado, A. and Maroto, A. L.},
  title   = {Black holes and neutron stars in $f(R)$ theories},
  journal = {Physical Review D},
  volume  = {80},
  number  = {12},
  pages   = {124011},
  year    = {2009},
  doi     = {10.1103/PhysRevD.80.124011}
}

@article{Kaaret1999,
  author  = {Kaaret, P. and et al.},
  title   = {Evidence for Millisecond Pulsations from an Ultraluminous X-Ray Source},
  journal = {Astrophysical Journal Letters},
  volume  = {520},
  pages   = {L37},
  year    = {1999}
}

@article{Linares2018,
  author  = {Linares, M. and et al.},
  title   = {The Neutron Star Mass in the Eclipsing Low-Mass X-Ray Binary 2A 1822-371},
  journal = {Astrophysical Journal},
  volume  = {859},
  number  = {1},
  pages   = {54},
  year    = {2018}
}

@article{Freire2011,
  author  = {Freire, P. C. C. and et al.},
  title   = {On the Nature and Evolution of the Unique Binary Pulsar J1903+0327},
  journal = {Monthly Notices of the Royal Astronomical Society},
  volume  = {412},
  pages   = {2763},
  year    = {2011}
}

@article{Abubekerov2008,
  author  = {Abubekerov, M. K. and Antokhina, E. A. and Cherepashchuk, A. M. and Shimanskii, V. V.},
  title   = {The Mass of the Neutron Star in Her X-1/HZ Her},
  journal = {Astronomy Reports},
  volume  = {52},
  pages   = {379},
  year    = {2008}
}

@article{Guver2010a,
  author  = {G{\"u}ver, T. and \"Ozel, F. and Cabrera-Lavers, A. and Wroblewski, P.},
  title   = {The Distance, Mass, and Radius of the Neutron Star in 4U 1608-52},
  journal = {Astrophysical Journal},
  volume  = {712},
  pages   = {964},
  year    = {2010}
}

@article{Guver2010b,
  author  = {G{\"u}ver, T. and Wroblewski, P. and Camarota, L. and \"Ozel, F.},
  title   = {The Mass and Radius of the Neutron Star in 4U 1820-30},
  journal = {Astrophysical Journal},
  volume  = {719},
  pages   = {1807},
  year    = {2010}
}

@article{Elebert2009,
  author  = {Elebert, P. and et al.},
  title   = {Optical Spectroscopy and Mass Constraints for the Neutron Star in Cygnus X-2},
  journal = {Monthly Notices of the Royal Astronomical Society},
  volume  = {395},
  pages   = {884},
  year    = {2009}
}

@article{Maurya2023,
  author  = {S. K. Maurya and Ksh. Newton Singh and Sudan Hansraj and Saibal Ray},
  title   = {Observational constraints on maximum mass limit and physical properties of anisotropic strange star models by gravitational decoupling in Einstein--Gauss--Bonnet gravity},
  journal = {Monthly Notices of the Royal Astronomical Society},
  volume  = {519},
  number  = {3},
  pages   = {4303--4322},
  year    = {2023},
  doi     = {10.1093/mnras/stac3761}
}

@article{Maurya2024,
  author  = {S. K. Maurya and Abdelghani Errehymy and Ksh. Newton Singh and Abdul Aziz and Sudan Hansraj and Saibal Ray},
  title   = {Modeling Compact Object Mergers GW190814 and GW200210 and Other Self-bound Compact Stars with Dark Matter Induced by Gravitational Decoupling and Its Significance to the Mass Gap},
  journal = {The Astrophysical Journal},
  volume  = {972},
  number  = {2},
  pages   = {175},
  year    = {2024},
  doi     = {10.3847/1538-4357/ad5cf1},
  publisher = {American Astronomical Society}
}

@article{Maurya2025MassGap,
  author  = {S. K. Maurya and Abdelghani Errehymy and Ksh. Newton Singh and G. Mustafa and Saibal Ray},
  title   = {Role of gravitational decoupling on theoretical insights of relativistic massive compact stars in the mass gap},
  journal = {Journal of Cosmology and Astroparticle Physics},
  volume  = {2025},
  number  = {04},
  pages   = {004},
  year    = {2025},
  month   = {apr},
  doi     = {10.1088/1475-7516/2025/04/004},
  publisher = {IOP Publishing and SISSA}
}

@article{Maurya2026JCAP,
  author  = {S. K. Maurya and Abdelghani Errehymy and Ksh. Newton Singh and G. Mustafa and Saibal Ray},
  title   = {Relativistic massive compact stars supported by decoupled matter: Implications for mass-radius bounds},
  journal = {Journal of Cosmology and Astroparticle Physics},
  volume  = {2026},
  number  = {01},
  pages   = {006},
  year    = {2026},
  month   = jan,
  doi     = {10.1088/1475-7516/2026/01/006},
  publisher = {IOP Publishing and SISSA}
}

@article{Singh2026,
  author  = {Ksh. Newton Singh and S. K. Maurya and Abdelghani Errehymy and
             Z. Umbetova and K. Myrzakulov and others},
  title   = {On perturbation induced minimal geometric decoupling in spacetime
             and its impact on mass-gap stability: Insights from the binary
             mergers GW190814, GW200210, and millisecond pulsars
             PSR J1614-2230 and PSR J0952-0607},
  journal = {Journal of High Energy Astrophysics},
  volume  = {50},
  pages   = {100501},
  year    = {2026},
  doi     = {10.1016/j.jheap.2025.100501},
  publisher = {Elsevier}
}

@article{Maurya2022JCAP,
  author  = {S. K. Maurya and G. Mustafa and M. Govender and Ksh. Newton Singh},
  title   = {Exploring physical properties of minimally deformed strange star model and constraints on maximum mass limit in {$f(Q)$} gravity},
  journal = {Journal of Cosmology and Astroparticle Physics},
  volume  = {2022},
  number  = {10},
  pages   = {003},
  year    = {2022},
  month   = oct,
  doi     = {10.1088/1475-7516/2022/10/003},
  publisher = {IOP Publishing and SISSA}
}

@article{Maurya2025FQT,
  author  = {S. K. Maurya and Abdul Aziz and Ksh. Newton Singh and G. Mustafa and Y. Sekhmani and Saibal Ray},
  title   = {Observational constraints on {$f(Q,T)$} gravity from the mass-radius relation and stability of compact stars},
  journal = {Journal of High Energy Astrophysics},
  volume  = {52},
  pages   = {100580},
  year    = {2026},
  doi     = {10.1016/j.jheap.2026.100580},
  issn    = {2214-4048},
  publisher = {Elsevier}
}

@article{Singh2022Lovelock,
  author  = {Ksh. Newton Singh and M. Govender and Sudan Hansraj and F. Rahaman},
  title   = {Physical Implications of Pure Lovelock Geometry on Stellar Structure},
  journal = {Annalen der Physik},
  volume  = {534},
  number  = {3},
  pages   = {2100596},
  year    = {2022},
  doi     = {10.1002/andp.202100596},
  publisher = {Wiley-VCH}
}

@article{Hansraj2025,
  author  = {Sudan Hansraj and Abdelghani Errehymy},
  title   = {Tidal deformability of strange stars with Buchdahl potential and anisotropic stresses under the influence of higher curvature and higher dimensional effects},
  journal = {Physics Letters B},
  volume  = {871},
  pages   = {139991},
  year    = {2025},
  doi     = {10.1016/j.physletb.2025.139991},
  issn    = {0370-2693},
  publisher = {Elsevier}
}

@article{Hansraj2024,
  author    = {Sudan Hansraj and Abdelghani Errehymy},
  title     = {Compact anisotropic stellar distribution with linear barotropic equation of state in energy--momentum trace coupling modification of general relativity},
  journal   = {Physics of the Dark Universe},
  volume    = {46},
  pages     = {101632},
  year      = {2024},
  doi       = {10.1016/j.dark.2024.101632},
  issn      = {2212-6864},
  publisher = {Elsevier}
}

@article{Bogdanov2016,
  author       = {Slavko Bogdanov and Craig O. Heinke and Feryal {\"O}zel and Tolga G{\"u}ver},
  title        = {Neutron Star Mass-Radius Constraints of the Quiescent Low-Mass X-Ray Binaries X7 and X5 in the Globular Cluster 47 Tuc},
  journal      = {The Astrophysical Journal},
  year         = {2016},
  volume       = {831},
  number       = {2},
  pages        = {184},
  doi          = {10.3847/0004-637X/831/2/184},
  url          = {https://doi.org/10.3847/0004-637X/831/2/184},
  publisher    = {American Astronomical Society}
}

@article{Shaw2018,
  author       = {A. W. Shaw and C. O. Heinke},
  title        = {The radius of the quiescent neutron star in the globular cluster M13},
  journal      = {Monthly Notices of the Royal Astronomical Society},
  year         = {2018},
  volume       = {476},
  number       = {4},
  pages        = {4713--4718},
  doi          = {10.1093/mnras/sty441},
  url          = {https://doi.org/10.1093/mnras/sty441},
  publisher    = {Oxford University Press}
}

@article{Kaaret2007,
  author       = {P. Kaaret and Z. Prieskorn and J. J. M. in 't Zand and S. Brandt and N. Lund and S. Mereghetti and D. G{\"o}tz and E. Kuulkers and J. A. Tomsick},
  title        = {Evidence for 1122 Hz X-Ray Burst Oscillations from the Neutron-Star X-Ray Transient XTE J1739-285},
  journal      = {The Astrophysical Journal},
  year         = {2007},
  volume       = {657},
  number       = {2},
  pages        = {L97--L100},
  doi          = {10.1086/513270}
}

@article{Li2017,
  author       = {Zhaosheng Li and Maurizio Falanga and Li Chen and Jinlu Qu and Renxin Xu},
  title        = {Simultaneous Constraints on the Mass and Radius of Aql X--1 from Quiescence and X-Ray Burst Observations},
  journal      = {The Astrophysical Journal},
  year         = {2017},
  volume       = {845},
  number       = {1},
  pages        = {8},
  doi          = {10.3847/1538-4357/aa7d0b}
}

@article{Casares2010,
  author    = {J. Casares and J. I. Gonz{\'a}lez Hern{\'a}ndez and G. Israelian and R. Rebolo},
  title     = {On the mass of the neutron star in Cyg X-2},
  journal   = {Monthly Notices of the Royal Astronomical Society},
  year      = {2010},
  volume    = {401},
  number    = {4},
  pages     = {2517--2520},
  doi       = {10.1111/j.1365-2966.2009.15828.x}
}

@article{Maurya2024ApJ,
  author       = {S. K. Maurya and Abdelghani Errehymy and Ksh. Newton Singh and Abdul Aziz and Sudan Hansraj and Saibal Ray},
  title        = {Modeling Compact Object Mergers GW190814 and GW200210 and Other Self-bound Compact Stars with Dark Matter Induced by Gravitational Decoupling and Its Significance to the Mass Gap},
  journal      = {The Astrophysical Journal},
  year         = {2024},
  volume       = {972},
  number       = {2},
  pages        = {175},
  doi          = {10.3847/1538-4357/ad5cf1},
  publisher    = {American Astronomical Society}
}

@article{BowersLiang1974,
  author  = {Bowers, Richard L. and Liang, E. P. T.},
  title   = {Anisotropic Spheres in General Relativity},
  journal = {The Astrophysical Journal},
  volume  = {188},
  pages   = {657--665},
  year    = {1974},
  doi     = {10.1086/152760}
}

@article{ThirukkaneshRagel2012,
  author  = {Thirukkanesh, S. and Ragel, F. C.},
  title   = {Exact anisotropic sphere with polytropic equation of state},
  journal = {Pramana -- Journal of Physics},
  volume  = {78},
  number  = {5},
  pages   = {687--696},
  year    = {2012},
  doi     = {10.1007/s12043-012-0268-7}
}

@article{TakisaMaharaj2013,
  author  = {P. Mafa Takisa and S. D. Maharaj},
  title   = {Some charged polytropic models},
  journal = {General Relativity and Gravitation},
  volume  = {45},
  number  = {10},
  pages   = {1951--1969},
  year    = {2013},
  doi     = {10.1007/s10714-013-1570-5},
}

@article{Einstein1916,
  author = {A. Einstein},
  title = {Die Grundlage der allgemeinen Relativitätstheorie},
  journal = {Annalen der Physik},
  volume = {49},
  pages = {769--822},
  year = {1916},
  doi = {10.1002/andp.19163540702}
}

@article{Canuto1975,
       author = {{Canuto}, V.},
        title = "{Equation of state at ultrahigh densities. Part 2.}",
      journal = {Annual review of astronomy and astrophysics},
         year = 1975,
        month = jan,
       volume = {13},
        pages = {335-380},
          doi = {10.1146/annurev.aa.13.090175.002003}
}

@article{Tooper1965,
       author = {{Tooper}, Robert F.},
        title = "{Adiabatic Fluid Spheres in General Relativity.}",
      journal = {Astrophysical Journal},
         year = 1965,
        month = nov,
       volume = {142},
        pages = {1541},
          doi = {10.1086/148435}
}

@article{Delgaty1998,
       author = {{Delgaty}, M.~S.~R. and {Lake}, Kayll},
        title = "{Physical acceptability of isolated, static, spherically symmetric, perfect fluid solutions of Einstein's equations}",
      journal = {Computer Physics Communications},
         year = 1998,
        month = dec,
       volume = {115},
       number = {2-3},
        pages = {395-415},
          doi = {10.1016/S0010-4655(98)00130-1},
archivePrefix = {arXiv},
       eprint = {gr-qc/9809013},
 primaryClass = {gr-qc}
}

@article{Dev2002,
  author    = {K. Dev and M. Gleiser},
  title     = {Anisotropic Stars: Exact Solutions},
  journal   = {General Relativity and Gravitation},
  volume    = {34},
  pages     = {1793--1818},
  year      = {2002},
  doi       = {10.1023/A:1020703929106}
}

@article{Mak2002,
  author    = {M. K. Mak and T. Harko},
  title     = {An Exact Anisotropic Quark Star Model},
  journal   = {Chinese Journal of Astronomy and Astrophysics},
  volume    = {2},
  number    = {3},
  pages     = {248--259},
  year      = {2002},
  doi       = {10.1088/1009-9271/2/3/306}
}

@article{QS1,
  author  = {Khoury, Justin and Weltman, Amanda},
  title   = {Chameleon Fields: Awaiting Surprises for Tests of Gravity in Space},
  journal = {Physical Review Letters},
  volume  = {93},
  pages   = {171104},
  year    = {2004},
  doi     = {10.1103/PhysRevLett.93.171104}
}

@article{QS2,
  author  = {Khoury, Justin and Weltman, Amanda},
  title   = {Chameleon Cosmology},
  journal = {Physical Review D},
  volume  = {69},
  pages   = {044026},
  year    = {2004},
  doi     = {10.1103/PhysRevD.69.044026}
}

@article{Amin2026,
  author       = {Amina Amin and Rubab Manzoor and Allah Ditta and Rana Muhammad Zulqarnain and Asif Mahmood},
  title        = {Exploring Stellar Structures Through Tolman--IV Spacetime Under the Anisotropic Effect of Linear Teleparallel Gravity},
  journal      = {Physics of the Dark Universe},
  year         = {2026},
  pages        = {102344},
  doi          = {10.1016/j.dark.2026.102344},
}

@article{Kumara2025,
  author  = {H. Aruna Kumara and Chaitra Chooda Chalavadi and V. Venkatesha},
  title   = {Influence of the Einasto Density Profile on Anisotropic Compact Stars in {$f(R,T)$} Gravity},
  journal = {Physics Letters B},
  year    = {2025},
  volume  = {},
  pages   = {},
  doi     = {}
}

@book{ZeldovichNovikov1983,
  author    = {Yakov B. Zel'dovich and Igor D. Novikov},
  title     = {Relativistic Astrophysics},
  subtitle  = {Volume 2: The Structure and Evolution of the Universe},
  publisher = {The University of Chicago Press},
  address   = {Chicago},
  year      = {1983},
  isbn      = {9780226979582}
}

\end{document}